\documentclass[12pt]{article}

\usepackage{jheppub, bm, wrapfig,float,array}
\usepackage[utf8]{inputenc}
\numberwithin{equation}{section}
\renewcommand\afterTocRuleSpace{\bigskip}
\usepackage{amsfonts}
\usepackage{amsmath}
\usepackage{color}
\usepackage{graphicx}
\usepackage{float}
\usepackage[T1]{fontenc}
\usepackage[utf8]{inputenc}
\usepackage{alphabeta}
\usepackage{fancyvrb}
\usepackage[dvipsnames]{xcolor}
\usepackage{bold-extra}
\usepackage{lmodern}

\usepackage{tikz}
\usetikzlibrary{arrows}
\usetikzlibrary{shapes.geometric,calc,arrows, positioning,shapes.misc,decorations.markings}
\tikzset{
  big arrow/.style={
    decoration={markings,mark=at position 1 with {\arrow[scale=2,#1]{>}}},
    postaction={decorate},
    shorten >=0.4pt},
  big arrow/.default=black}

\newcommand{\bea}{\begin{eqnarray}}
\newcommand{\eea}{\end{eqnarray}}
\newcommand{\be}{\begin{equation}}
\newcommand{\ee}{\end{equation}}
\newcommand{\bit}{\begin{itemize}}
\newcommand{\eit}{\end{itemize}}
\newcommand{\ben}{\begin{enumerate}}
\newcommand{\een}{\end{enumerate}}

\renewcommand{\ni}{\noindent}

\newcommand{\half}{\frac{1}{2}}

\newcommand{\Z}{{\mathbb Z}}

\newcommand{\C}{{\mathbb C}}

\renewcommand{\P}{{\mathbb P}}

\newcommand{\cC}{\mathcal{C}}

\newcommand{\cN}{\mathcal{N}}
\newcommand{\cO}{\mathcal{O}}

\newcommand{\cR}{\mathcal{R}}

\newcommand{\cT}{\mathcal{T}}

\newcommand{\F}{\mathsf{F}}
\renewcommand{\S}{\mathsf{S}}

\renewcommand{\C}{\mathsf{C}}
\newcommand{\A}{\mathsf{A}}

\renewcommand{\C}{\mathsf{C}}

\renewcommand{\L}{\mathsf{\Lambda}}

\newcommand{\fT}{\mathfrak{T}}

\newcommand{\fe}{\mathfrak{e}}
\newcommand{\ff}{\mathfrak{f}}
\newcommand{\fg}{\mathfrak{g}}
\newcommand{\fh}{\mathfrak{h}}
\newcommand{\su}{\mathfrak{su}}
\renewcommand{\sp}{\mathfrak{sp}}
\newcommand{\so}{\mathfrak{so}}
\renewcommand{\u}{\mathfrak{u}}

\newcommand{\ubf}[1]{\underline{\bf #1}}

\newcommand{\bF}{{\mathbb F}}

\tikzstyle{brane}=[draw]
\tikzset{D7/.style={circle, draw=black, inner sep=0pt, fill=white, minimum size=3mm}}
\tikzset{hasse/.style={circle, fill,inner sep=2pt}}
\tikzset{flavor/.style={regular polygon,regular polygon sides=4,inner sep=2.5pt, draw}}
\tikzset{gauge/.style={circle, draw,inner sep=2.5pt}}
\tikzset{gaugeb/.style={circle, draw,fill=black,inner sep=2.5pt}}
\tikzset{gaugecyan/.style={circle, draw,fill=cyan,inner sep=2.5pt}}
\tikzset{gaugegreen/.style={circle, draw,fill=green,inner sep=2.5pt}}
\tikzset{gaugeblue/.style={circle, draw,fill=blue,inner sep=2.5pt}}
\tikzset{gaugeorange/.style={circle, draw,fill=orange,inner sep=2.5pt}}
\tikzset{bd/.style={circle, draw=black, inner sep=0pt, fill=black, minimum size=2mm}}
\tikzset{wd/.style={circle, draw=black, inner sep=0pt, fill=white, minimum size=2mm}}
\tikzset{Dynkin/.style={circle, draw=black, inner sep=0pt, fill=white, minimum size=2mm}}
\tikzstyle{ligne}=[draw, thick] 
\tikzset{doublearrow/.style={ draw=black!75, color=black!75, thick, double distance=3pt, }}

\title{Higher-form symmetries of $6d$ and $5d$ theories}

\author{Lakshya Bhardwaj$\,^a$, Sakura Sch\"{a}fer-Nameki$\,^{b}$}

\affiliation{$^a$ Department of Physics, Harvard University\\ \hspace*{8pt}17 Oxford St, Cambridge, MA 02138, USA\\
$^b$ Mathematical Institute, University of Oxford,\\
\phantom{$^b$} Andrew Wiles Building, Woodstock Road, Oxford, OX2 6GG, UK}

\abstract{We describe general methods for determining higher-form symmetry groups of known $5d$ and $6d$ superconformal field theories (SCFTs), and $6d$ little string theories (LSTs). The $6d$ theories can be described as supersymmetric gauge theories in $6d$ which include both ordinary non-abelian 1-form gauge fields and also abelian 2-form gauge fields. Similarly, the $5d$ theories can also be often described as supersymmetric non-abelian gauge theories in $5d$. Naively, the 1-form symmetry of these $6d$ and $5d$ theories is captured by those elements of the center of ordinary gauge group which leave the matter content of the gauge theory invariant. However, an interesting subtlety is presented by the fact that some massive BPS excitations, which includes the BPS instantons, are charged under the center of the gauge group, thus resulting in a further reduction of the 1-form symmetry. We use the geometric construction of these theories in M/F-theory to determine the charges of these BPS excitations under the center. We also provide an independent algorithm for the determination of 1-form symmetry for $5d$ theories that admit a generalized toric construction (i.e. a 5-brane web construction). The 2-form symmetry group of $6d$ theories, on the other hand, is captured by those elements of the center of the abelian 2-form gauge group that leave all the massive BPS string excitations invariant, which is much more straightforward to compute as it is encoded in the Green-Schwarz coupling associated to the $6d$ theory.
}

\begin{document}

\maketitle

\section{Introduction}

Higher-form global symmetries \cite{Gaiotto:2014kfa} of theories play an important role in characterizing refined properties, such as the spectrum of line- and higher-dimensional defect operators. In the simplest instance they correspond to the center symmetries of Yang-Mills theories, under which the Wilson lines are charged. 
In higher dimensions, in particular 5d and 6d much recent progress has been made in uncovering properties of superconformal field theories (SCFTs) and related theories, such as little string theories (LSTs). SCFTs in 5d and 6d are intrinsically strongly coupled, and have an IR description in terms of an effective theory on the Coulomb branch and tensor branch, respectively. One of the questions that we will address in this paper is how to determine the higher-form symmetries of the quantum theories from the effective description. The key subtlety here is the existence of instanton particles or strings, which can be charged under the one-form symmetry, and can thereby break the symmetry. 

This will be complemented with the analysis in geometry, using either the description in terms of collapsable surfaces or a description in terms of generalized toric diagrams (i.e. 5-brane-webs). Much progress has been made on mapping out the theories in 6d, including a putative classification of SCFTs \cite{Heckman:2013pva, Heckman:2015bfa, Bhardwaj:2015xxa, Bhardwaj:2019hhd} and LSTs \cite{Bhardwaj:2015oru, Bhardwaj:2019hhd} from F-theory on elliptic Calabi-Yau three-folds -- for a review of the 6d classification, see \cite{Heckman:2018jxk}. In 5d recent progress has been made in mapping out and furthering the classification of SCFTs using the M-theory realization on canonical singularities  \cite{Hayashi:2014kca, DelZotto:2017pti, Jefferson:2017ahm, Closset:2018bjz, Jefferson:2018irk, Apruzzi:2018nre, Bhardwaj:2018yhy, Bhardwaj:2018vuu, Apruzzi:2019vpe, Apruzzi:2019opn, Apruzzi:2019enx, Bhardwaj:2019jtr, Apruzzi:2019kgb, Bhardwaj:2019fzv, Bhardwaj:2019xeg, Eckhard:2020jyr, Bhardwaj:2020kim, Closset:2020scj}. 

Higher form symmetries in 6d and 5d SCFTs are highly constrained by the superconformal algebra. 
As is shown in \cite{Cordova:2016emh} (and related upcoming work by the same authors), there cannot be any continous 1-form symmetry in such theories. Indeed, we will see that 1-form symmetries 5d and 6d SCFTs are discrete. 
From a geometric engineering point of view, higher form symmetries were discussed using the M-theory realization of 5d SCFTs on Calabi-Yau threefolds, as well as other M-theory geometric engineering constructions such as $G_2$-holonomy compactifications to 4d in \cite{Morrison:2020ool, Albertini:2020mdx, Eckhard:2020jyr}. Related works in Type IIB, for 4d SCFTs in particular Argyres-Douglas theories were obtained in \cite{Garcia-Etxebarria:2019cnb, Closset:2020scj, DelZotto:2020esg}. In 6d the defect group was analyzed in \cite{DelZotto:2015isa} and the 1-form symmetries in 6d SCFTs were discussed from a geometric construction in \cite{Morrison:2020ool}. The global form of the flavor symmetry in 6d was discussed in \cite{Dierigl:2020myk}, using the torsional part of the Mordell-Weil group of elliptic fibrations in F-theory. 
In this paper the main new insight is two-fold: we determine how to compute the higher form symmetry from the effective description in the IR, taking into account non-perturbative instanton effects. We observe that in many cases these non-perturbative effects are correlated with the existence of half-hypers in the theory, i.e. if the half-hypers are completed into full hypers, the non-perturbative effects disappear. The other aspect of this paper is the generalization to arbitrary 6d and 5d theories. This includes 6d SCFTs, LSTs and the frozen phases of F-theory \cite{Tachikawa:2015wka, Bhardwaj:2018jgp}. 
6d theories are closely connected with 5d theories by circle-reduction, with potentially added holonomies (in flavor symmetries), or twists. We track the higher form symmetries through this dimensional reduction and match it with one computed in 5d. This provides another confirmation for the approach we propose, and confirms the geometric analysis. 

In 5d a complementary approach uses the 5-brane webs, which engineer a class of 5d SCFTs. These are dual to so-called generalized toric polygons (or dot diagrams) \cite{Benini:2009gi}. We provide a prescription generalizing the analysis for toric models for computing the 1-form symmetry for generalized toric polygons, and underpin this with a discussion of the Wilson lines in the 5-brane web. 

The plan of the paper is as follows: in section \ref{6} we discuss the 6d case, starting with the 2-form symmetry  in 6d SCFTs and LSTs, followed by their 1-form symmetry. In section \ref{5} the 5d theories are discussed, both in terms of their relation to 6d theories, and the analysis on the Coulomb branch. We furthermore provide an analysis of the 5d theories that have a description as brane-webs, or generalized toric diagrams.

\section{Higher-form symmetries of $6d$ SCFTs and LSTs}\label{6}

This section is devoted to the study of higher-form symmetries in supersymmetric $6d$ theories. There are two known kinds of UV complete theories in six dimensions which do not include dynamical gravity. The first are supersymmetric conformal field theories (SCFTs), and the second are supersymmetric little string theories (LSTs).

We would like to argue that it is sufficient for us to focus on a class of $6d$ theories\footnote{From this point onward, a ``$6d$ theory'' will refer to either a $6d$ SCFT or a $6d$ LST.} which admit only two different kinds of higher-form symmetry groups, namely discrete 1-form symmetry group $\cO$ and discrete 2-form symmetry group $\cT$. One can obtain theories outside this class by performing various kinds of discrete gaugings. For example, one can gauge a subgroup $\cO'$ of the 1-form symmetry $\cO$ to obtain a $6d$ theory with discrete 3-form symmetry group. One can also stack the $6d$ theory with an SPT phase carrying 1-form symmetry $\cO'$ before gauging the diagonal $\cO'$ symmetry, thus producing more $6d$ theories which have 3-form symmetries. It might also be possible to obtain $6d$ theories carrying 4-form symmetry by gauging discrete subgroups, possibly in combination with an SPT phase, of the 0-form symmetry group of the above special class of $6d$ theories. At the time of writing of this paper, there is no known $6d$ theory that cannot be produced as a discrete gauging of the above class of $6d$ theories. For any such discrete gauging, the spectrum of higher-form symmetries (along with possible higher-group structures) and their anomalies can  be deduced from the knowledge of the spectrum of higher-form symmetries and anomalies of the above special class of $6d$ theories.

Moreover, at the time of writing of this paper, all the known $6d$ theories in the above class admit a geometric construction in F-theory\footnote{These constructions can be divided into two types. The first kind of constructions are referred to be in the ``unfrozen phase'' of F-theory and do not involve O7$^+$ planes. The second type of constructions are referred to be in the ``frozen phase'' of F-theory \cite{Bhardwaj:2018jgp,Tachikawa:2015wka} and involve O7$^+$ planes. See \cite{Heckman:2015bfa,Bhardwaj:2015oru} for classification of theories of first type and \cite{Bhardwaj:2019hhd} for classification of theories of second type.}. In this paper, we thus focus only on the above set of ``F-theoretic'' $6d$ theories and provide methods to determine their 1-form and 2-form symmetry groups.

Our analysis will involve passing on to a generic point on the tensor branch of vacua\footnote{Note that every known F-theoretic $6d$ theory admits a tensor branch of vacua.} of the $6d$ theory. We will assume that the full higher-form symmetry of the $6d$ theory is visible at such a point on the tensor branch, if we also take into account massive BPS excitations in the theory on the tensor branch. We will be presenting our analysis in field-theory terms without referring to the technicalities of F-theory construction. An advantage of this approach is that it allows us to treat the $6d$ theories arising from both the unfrozen and the frozen phases of F-theory on an equal footing.

At a generic point on the tensor branch, an F-theoretic $6d$ SCFT or LST flows to a $6d$ $\cN=(1,0)$ gauge theory (carrying a semi-simple gauge algebra) along with a set of free tensor multiplets\footnote{For an SCFT all these tensor multiplets are dynamical, while for an LST one of the tensor multiplets is a non-dynamical background tensor multiplet.} in the IR. Moreover, the theory on the tensor branch carries massive BPS string excitations in one-to-one correspondence with a special basis for these tensor multiplets. These strings are charged under the 2-form gauge fields living in the tensor multiplets. Their charges are captured by a symmetric positive semi-definite integer matrix $\Omega^{ij}$ (which is the matrix participating in Green-Schwarz mechanism of gauge anomaly cancellation) with non-positive off-diagonal entries, where $i$ labels different elements in the above-mentioned special basis for the tensor multiplets. This matrix $\Omega^{ij}$ is positive definite for a $6d$ SCFT, and it is a positive semi-definite matrix of corank 1 for an irreducible\footnote{We call an LST irreducible if it cannot be written as a stack product of other LSTs.} LST. The rank of $\Omega^{ij}$ will be denoted by $r$ in what follows, and it is also known as the rank of the $6d$ SCFT or LST to which $\Omega^{ij}$ is associated.

A subset of the above mentioned BPS strings arise as the BPS instanton strings for the simple factors in the low-energy gauge algebra. Thus, each simple factor of the gauge algebra is associated to some $i$ and we refer to the corresponding simple factor of gauge algebra as $\fg_i$.

We can thus denote a $6d$ SCFT or LST by displaying the above discussed data in a graphical notation of the following form:
\be\label{TBD}
\begin{tikzpicture}
\node (v1) at (-0.5,0.5) {$\Omega^{ii}$};
\node (v4) at (-0.5,1) {$\fg_i$};
\begin{scope}[shift={(2.8,0.05)}]
\node at (-0.5,0.9) {$\fg_j$};
\node (v2) at (-0.5,0.45) {$\Omega^{jj}$};
\end{scope}
\node (v3) at (0.9,0.5) {\tiny{$-\Omega^{ij}$}};
\draw (v1)--(v3);
\draw (v2)--(v3);
\begin{scope}[shift={(-2.3,0.05)}]
\node (v2_1) at (-0.5,0.45) {$\Omega^{kk}$};
\end{scope}
\begin{scope}[shift={(0,1.95)}]
\node (v2_2) at (-0.5,0.45) {$\Omega^{ll}$};
\node at (-0.5,0.9) {$\fg_l$};
\end{scope}
\draw  (v2_2) edge (v4);
\draw  (v2_1) edge (v1);
\end{tikzpicture} \,,
\ee
where there is a node for each $i$. Each node is labeled by $\Omega^{ii}$ and the associated gauge algebra $\fg_i$. We leave $\fg_i$ empty for a node $i$ if the BPS string corresponding to that node is not an instanton string of any gauge algebra. The node labeled as $k$ in the above graph is such an example. Two nodes $i$ and $j$ are connected by an edge if the off-diagonal entry $\Omega^{ij}\neq0$. If furthermore $-\Omega^{ij}>1$, then we insert a label at the middle of the edge indicating this number $-\Omega^{ij}$. If $-\Omega^{ij}=1$, then no such label is inserted. The edge between $i$ and $l$ in the above graph is such an example. See \cite{Bhardwaj:2019fzv} for more details on this notation in the context of $6d$ SCFTs.

\subsection{2-form symmetry}\label{6T}
\subsubsection{2-form symmetry of $6d$ SCFTs}
If we forget about the BPS strings for a moment, then there is a $U(1)$ 2-form symmetry associated to each tensor multiplet $i$ under which the ``Wilson surface'' for the 2-form gauge field living within the tensor multiplet $i$ has charge 1. Thus, we obtain a potential $U(1)^r$ 2-form symmetry\footnote{It should be noted that this $U(1)^r$ 2-form symmetry is spontaneously broken along the tensor branch. This is akin to the spontaneous breaking of $U(1)^r$ electric 1-form symmetry in an abelian gauge theory \cite{Gaiotto:2014kfa}. Since the 2-form symmetry $\cT$ of $6d$ SCFTs and LSTs embeds into this $U(1)^r$ 2-form symmetry, $\cT$ is always spontaneously broken along the tensor branch as well. We expect $\fT$ to be spontaneously broken at the conformal point of a $6d$ SCFT too.}. When the BPS strings are included, the 2-form symmetry is reduced to the subgroup of $U(1)^r$ under which the BPS strings are uncharged.

The 2-form symmetry in the presence of the charged strings is then found by computing the Smith normal form $T^{ij}$ of the matrix $\Omega^{ij}$, which, due to the positive definiteness of $\Omega^{ij}$, is a diagonal matrix with the diagonal entries being positive integers. Let $n_i$ be the $i$-th diagonal entry of $T^{ij}$. Then the 2-form symmetry group $\cT$ can be written as
\be\label{2g}
\cT=\prod_{i=1}^r~\Z_{n_i}\,,
\ee
i.e. a product of $\Z_{n_i}$ for all $i$, where $\Z_1$ denotes the trivial group.

The appearance of the Smith normal form is easy to understand from the point of view of Pontryagin dual of the 2-form symmetry group. Before accounting for the charged strings, the dual is the lattice $\Z^r$ which captures the possible charges of surface defects and dynamical strings under the 2-form gauge fields. The matrix $\Omega^{ij}$ defines a sublattice $[\Omega^{ij}]\cdot \Z^r$ inside the lattice $\Z^r$ which is spanned by vectors
\be
v^i:=\sum _j\Omega^{ij}u_j \,,
\ee
where $u_i$ is the standard basis of $\Z^r$. This sublattice captures the charges of the dynamical strings. The charges under $\cT$ are then captured by the quotient lattice 
\be
\frac{\Z^r}{[\Omega^{ij}]\cdot \Z^{r}}\,,
\ee
whose Pontryagin dual is $\cT$. After changing the basis inside $\Z^r$ and $[\Omega^{ij}]\cdot \Z^r$, we can write the above quotient lattice as
\be
\frac{\Z^r}{[T^{ij}]\cdot \Z^{r}}=\bigoplus_{i=1}^r~\frac{\Z}{n_i\Z}\,,
\ee
The Pontryagin dual of each subfactor is isomorphic to itself since $n_i>0$, and hence we find that the 2-form symmetry group $\cT$ is as shown in (\ref{2g}).

\subsubsection{2-form symmetry of $6d$ LSTs}
The structure of $6d$ LSTs is similar to that of $6d$ SCFTs, the crucial difference being that the matrix $\Omega^{ij}$ is only positive semi-definite for $6d$ LSTs. Naively, one might expect that the 2-form symmetry group for an LST would be captured by the quotient lattice
\be
\frac{\Z^{r+1}}{[\Omega^{ij}]\cdot \Z^{r+1}}\,,
\ee
where the total number of nodes $i$ is $r+1$ as $\Omega^{ij}$ has rank $r$ and corank 1 for an irreducible LST. The fact that the corank of $\Omega^{ij}$ is 1 implies that the above quotient lattice contains one factor of $\Z$ along with a torsion part. That is, the above quotient lattice takes the following form
\be
\bigoplus_{i=1}^r~\frac{\Z}{n_i\Z}\oplus\Z\,.
\ee
Taking its Pontryagin dual, the above naive expectation would lead us to believe that the 2-form symmetry group for a LST takes the form
\be\label{6L}
\prod_{i=1}^r~\Z_{n_i}\times U(1)\,.
\ee
However, we must take into account the fact that one of the tensor multiplets, out of the $r+1$ tensor multiplets associated to the nodes $i$, is non-dynamical. Hence this tensor multiplet does not generate a potential $U(1)$ 2-form symmetry, and we should mod out this $U(1)$ factor from (\ref{6L}) since we have taken it into account in our above calculation. Thus, the 2-form symmetry of a little string theory is
\be
\cT=\prod_{i=1}^r~\Z_{n_i}\,.
\ee

\subsubsection{Examples}
\ubf{Example 1}: Consider the case of $\cN=(2,0)$ SCFTs. These can be described in terms of a simply laced simple Lie algebra $\fg$. The matrix $\Omega^{ij}$ is the Cartan matrix of $\fg$. Then, $\cT$ simply coincides with the center of $\fg$.

Similarly, $\cN=(2,0)$ LSTs are also described in terms of a simply laced simple Lie algebra $\fg$ but the associated matrix $\Omega^{ij}$ is the Cartan matrix of $\fg^{(1)}$, which is the untwisted affine Lie algebra associated to $\fg$. Again, $\cT$ coincides with the center of $\fg$.

\vspace{8pt}

\ni\ubf{Example 2}: Consider the following $6d$ SCFT arising in the \emph{frozen} phase of F-theory
\be
\begin{tikzpicture}
\node (v1) at (-0.5,0.5) {$4$};
\node (v4) at (-0.5,1) {$\so(n)$};
\begin{scope}[shift={(2.2,0.05)}]
\node (v2) at (-0.5,0.45) {2};
\end{scope}
\node (v3) at (0.6,0.5) {\tiny{2}};
\draw (v1)--(v3);
\draw (v2)--(v3);
\node (v4) at (1.7,1) {$\su(n-8)$};
\end{tikzpicture}
\ee
Its associated tensor branch gauge theory contains gauge algebra $\so(n)\oplus\su(n-8)$ with the matter content being a hyper in bifundamental representation plus $n-16$ hypers in fundamental representation of $\su(n-8)$. The matrix $\Omega^{ij}$ for this theory is
\be
\begin{pmatrix}
4&-2\\
-2&2
\end{pmatrix} \,.
\ee
The Smith normal form of the above matrix is
\be
\begin{pmatrix}
2&\:\:\:\:\:\:\\
\:\:\:\:\:\:&2
\end{pmatrix}\,,
\ee
and thus $\cT=\Z_2\times\Z_2$.

\vspace{8pt}

\ni\ubf{Example 3}: Consider the LST
\be
\begin{tikzpicture}
\node (v1) at (-0.5,0.5) {$4$};
\node (v4) at (-0.5,1) {$\so(2n+8)$};
\begin{scope}[shift={(2.2,0.05)}]
\node (v2) at (-0.5,0.45) {1};
\end{scope}
\node (v3) at (0.6,0.5) {\tiny{2}};
\draw (v1)--(v3);
\draw (v2)--(v3);
\node (v4) at (1.7,1) {$\sp(n)$};
\end{tikzpicture}
\ee
whose tensor branch gauge theory contains a \emph{full} hypermultiplet in the bifundamental. The 2-form symmetry group can be computed to be
\be
\cT=\Z_1 \,.
\ee

One can obtain a $6d$ SCFT from a LST by deleting a node. Note that a $6d$ SCFT obtained this way need not have the same 2-form symmetry group as that of the $6d$ LST. For example, deleting the $\sp(n)$ node in the above LST, we obtain the $6d$ SCFT
\be
\begin{tikzpicture}
\node (v1) at (-0.5,0.5) {$4$};
\node (v4) at (-0.5,1) {$\so(2n+8)$};
\end{tikzpicture} 
\ee
for which $\cT=\Z_4$.

\subsubsection{Relative nature of $6d$ SCFTs and LSTs}
General $6d$ SCFTs and LSTs are relative theories, which means that they are more properly thought of as theories living on the boundaries of some particular kind of $7d$ topological quantum field theories (TQFTs). It is well-known in the context of $6d$ SCFTs having a construction in the unfrozen phase of F-theory that the $7d$ TQFT associated to such a $6d$ SCFT is captured by the 2-form symmetry group (also known as the \emph{defect group} \cite{DelZotto:2015isa}) $\cT$ of the $6d$ SCFT.

This should admit a straightforward generalization to $6d$ SCFTs constructed in the frozen phase of F-theory and LSTs, for which the recipe to compute $\cT$ has been provided above.

\subsection{1-form symmetry of $6d$ SCFTs and LSTs}\label{6O}
If we forget about the hypermultiplet matter content of the $\cN=(1,0)$ low-energy gauge theory and the dynamical BPS strings, then the 1-form symmetry is the product of the center\footnote{More precisely, we are working with a form of the theory where the gauge groups $G_i$ realizing all the gauge algebras $\fg_i$ are simply connected. Other forms of the theory having non-simply-connected gauge groups can be obtained from this form of the theory by gauging the 1-form symmetries, if any. Throughout this paper, we will abuse the language and refer to the center $Z(G)$ of the simply connected group $G$ of a simple algebra $\fg$ as the ``center of the simple algebra $\fg$''.} $\Gamma_i$ of each simple factor $\fg_i$ of the tensor branch gauge algebra\footnote{The $\cN=(1,0)$ low-energy non-abelian gauge theory is free in the extreme IR, and hence described by a bunch of free vector multiplets in the far IR. The 1-form symmetry associated to these free vector multiplets is spontaneously broken. Since, as we will see, the 1-form symmetry $\cO$ of the 6d SCFT or LST is a subgroup of $\prod \Gamma_i$ 1-form symmetry which is further embedded into the 1-form symmetry of the free vector multiplets in the IR, $\cO$ is spontaneously broken along the tensor branch. We also expect $\cO$ to be spontaneously broken at the conformal point of a $6d$ SCFT.}. Including the hypermultiplets and BPS strings, the 1-form symmetry $\cO$ of the theory becomes the subgroup of $\prod_i\Gamma_i$ under which all hypermultiplets and BPS strings are uncharged.

The charges of (full or half) hypermultiplets under $\prod_i\Gamma_i$ is determined by knowing the representation $R$ of $\fg=\oplus_i\fg_i$ formed by these hypermultiplets. We will describe a way to compute the charge of any arbitrary representation $R$ under $\prod_i\Gamma_i$ in Section \ref{5GG}. The charges of representations relevant in the context of $6d$ SCFTs and LSTs are displayed in Table \ref{table}. The charges for arbitrary reps are provided in equations (\ref{ch1}) and (\ref{ch2}).

\begin{table}[h]
\begin{center}
\begin{tabular}{ | l | c | c | c | }
\hline
Gauge algebra & Center & Representations & Charge \\ \hline\hline
$\su(n)$ & $\Z_n$ & \parbox[t]{0.35cm}{$\F$\\$\L^2$\\$\L^3$\\$\S^2$} & \parbox[t]{4.1cm}{$1~(\text{mod}~n)$\\$2~(\text{mod}~n)$\\$3~(\text{mod}~n)$\\$2~(\text{mod}~n)$} \\ \hline
$\so(2n+1)$ & $\Z_2$ & \parbox[t]{0.35cm}{$\F$\\$\S$} & \parbox[t]{4.1cm}{$0~(\text{mod}~2)$\\$1~(\text{mod}~2)$} \\ \hline
$\sp(n)$ & $\Z_2$ & \parbox[t]{0.35cm}{$\F$\\$\L^2$\\$\L^3$} & \parbox[t]{4.1cm}{$1~(\text{mod}~2)$\\$0~(\text{mod}~2)$\\$1~(\text{mod}~2)$} \\ \hline
$\so(4n+2)$ & $\Z_4$ & \parbox[t]{0.35cm}{$\F$\\$\S$\\$\C$} & \parbox[t]{4.1cm}{$2~(\text{mod}~4)$\\$1~(\text{mod}~4)$\\$3~(\text{mod}~4)$} \\ \hline
$\so(4n)$ & $\Z_2\times\Z_2$ & \parbox[t]{0.35cm}{$\F$\\$\S$\\$\C$} & \parbox[t]{4.1cm}{$\left(1~(\text{mod}~2),1~(\text{mod}~2)\right)$\\$\left(1~(\text{mod}~2),0~(\text{mod}~2)\right)$\\$\left(0~(\text{mod}~2),1~(\text{mod}~2)\right)$} \\ \hline
$\fe_6$ & $\Z_3$ & \parbox[t]{0.35cm}{$\F$} & \parbox[t]{4.1cm}{$1~(\text{mod}~3)$} \\ \hline
$\fe_7$ & $\Z_2$ & \parbox[t]{0.35cm}{$\F$} & \parbox[t]{4.1cm}{$1~(\text{mod}~2)$} \\ \hline
$\fe_8$ & $\Z_1$ & \parbox[t]{0.35cm}{$\F$} & \parbox[t]{4.1cm}{$0~(\text{mod}~1)$} \\ \hline
$\ff_4$ & $\Z_1$ & \parbox[t]{0.35cm}{$\F$} & \parbox[t]{4.1cm}{$0~(\text{mod}~1)$} \\ \hline
$\fg_2$ & $\Z_1$ & \parbox[t]{0.35cm}{$\F$} & \parbox[t]{4.1cm}{$0~(\text{mod}~1)$} \\ \hline
\end{tabular}
\end{center}
\caption{Centers of various gauge algebras and charges of some of the representations under the center of the gauge algebra. The adjoint representation $\A$ is not mentioned in the table above since it always has charge 0 under the corresponding center. $\F$ denotes the fundamental representation for $\su(n),\sp(n)$; the vector representation for $\so(n)$; and the irreducible representations of dimensions $\mathbf{7},\mathbf{26},\mathbf{27},\mathbf{56}$ for $\fg_2,\ff_4,\fe_6,\fe_7$ respectively. We often refer to $\F$ as the ``fundamental representation'' of the corresponding algebra. $\Lambda^2$ and $\Lambda^3$ denote the irreducible two and three index antisymmetric representations for $\su(n)$ and $\sp(n)$. $\S^2$ denotes the two-index symmetric irrep for $\su(n)$. $\S$ and $\C$ denote the irreducible spinor reps of different chirality for $\so(2n)$; and $\S$ denotes the irreducible spinor rep for $\so(2n+1)$. The charges for arbitrary irreps are provided in equations (\ref{ch1}) and (\ref{ch2}).}
\label{table}
\end{table}

As far as charges of BPS strings are concerned, it is often the case that the charges of BPS strings under $\prod_i\Gamma_i$ are already accounted by the charges of hypermultiplets under $\prod_i\Gamma_i$. However, in some cases, BPS strings lead to independent contributions not accounted by the hypermultiplets. The hallmark of these cases is that either they involve tensor multiplets that are not paired to a gauge algebra, or the matter content is such that we have a half-hyper in some irreducible representation of $\fg=\oplus_i\fg_i$, or the $\Z_2$ valued theta angle of a $\fg_i=\sp(n)$ is relevant. More exhaustively, these cases are listed below:
\ben
\item Consider a node $i$ with $\Omega^{ii}=1$ and $\fg_i$ trivial. Then, look at the set\footnote{This set is trivial if there is a node $j$ with $\Omega^{ij}<-1$. See the discussion later in this subsection accounting for the possibility of such nodes.} of nodes $j$ such that $\Omega^{ij}=-1$ and $\fg_j$ is non-trivial. It is well-known that the sum $\oplus_j\fg_j$ of these $\fg_j$ is a subalgebra of $\fe_8$. Correspondingly, the adjoint representation of $\fe_8$ decomposes as some representation $\cR$ of $\oplus_j\fg_j$. Then, the charge of the BPS string corresponding to node $i$ is captured by the charge of $\cR$ under $\prod_j\Gamma_j$.\\
Schematically the graph near the node $i$ takes the following form
\be
\begin{tikzpicture}
\node (v1) at (-0.5,0.5) {$1$};
\begin{scope}[shift={(2,0.05)}]
\node at (-0.5,0.9) {$\fg_j$};
\node (v2) at (-0.5,0.45) {$\Omega^{jj}$};
\end{scope}
\begin{scope}[shift={(-2,0.05)}]
\node (v2_1) at (-0.5,0.45) {$\Omega^{kk}$};
\node at (-0.5,0.9) {$\fg_k$};
\end{scope}
\begin{scope}[shift={(0,1.95)}]
\node (v2_2) at (-0.5,0.45) {$\Omega^{ll}$};
\node at (-0.5,0.9) {$\fg_l$};
\end{scope}
\draw  (v2_2) edge (v1);
\draw  (v2_1) edge (v1);
\draw  (v1) edge (v2);
\end{tikzpicture}
\ee
\item Consider a situation, where we have two nodes $i$ and $j$ such that $\fg_i=\sp(n)$ and $\fg_j=\so(m)$ for $n>0$ and $m\neq8$, and $\Omega^{ij}=-1$. The matter content between $\sp(n)$ and $\so(m)$ is a half-hyper in a mixed representation of $\sp(n)\oplus\so(m)$ with mixed representation being the bifundamental representation. In this case, we need to account for the charge of BPS instanton strings for $\sp(n)$ under center $\Gamma_j$ of $\so(m)$. We can take this string to be charged under $\Gamma_j$ as the irreducible spinor representation $\S$ of $\so(m)$ is charged under $\Gamma_j$.\\
Schematically the graph near the nodes $i$ and $j$ takes the following form
\be\label{sf}
\begin{tikzpicture}
\node (v1) at (-0.5,0.5) {$1$};
\node (v4) at (-0.5,1) {$\sp(n)$};
\begin{scope}[shift={(2,0.05)}]
\node at (-0.5,0.9) {$\so(m)$};
\node (v2) at (-0.5,0.45) {$\Omega^{jj}$};
\end{scope}
\begin{scope}[shift={(4.2,0.05)}]
\node (v2_1) at (-0.5,0.45) {$\Omega^{kk}$};
\end{scope}
\begin{scope}[shift={(-2.5,0.05)}]
\node (v2_2) at (-0.5,0.45) {$\Omega^{ll}$};
\node at (-0.5,0.9) {$\fg_l$};
\end{scope}
\draw  (v2_2) edge (v1);
\draw  (v2_1) edge (v2);
\begin{scope}[shift={(0,-1.7)}]
\node (v3) at (-0.5,0.9) {$\fg_m$};
\node (v2_3) at (-0.5,0.45) {$\Omega^{mm}$};
\end{scope}
\draw  (v1) edge (v2);
\draw  (v1) edge (v3);
\end{tikzpicture}
\ee
\item Now, consider a situation where we have two nodes $i$ and $j$ such that $\fg_i=\sp(n)$ and $\fg_j=\so(8)$ for $n>0$, and $\Omega^{ij}=-1$. In this case, the matter content between $\sp(n)$ and $\so(8)$ is a half-hyper in a mixed representation of $\sp(n)\oplus\so(8)$. The mixed representation takes the form $\F\otimes\cR$ where $\F$ is the fundamental representation of $\sp(n)$ and $\cR$ is one of the following 3 representations of $\so(8)$: vector $\F$, spinor $\S$, or cospinor $\C$. If $\cR=\F$, then the charge of BPS instanton string for $\sp(n)$ under $\Gamma_j$ can be taken to be the same as that of the representation $\S$ of $\so(8)$. If $\cR=\S$, then the charge of BPS instanton string for $\sp(n)$ under $\Gamma_j$ can be taken to be the same as that of the representation $\C$ of $\so(8)$. If $\cR=\C$, then the charge of BPS instanton string for $\sp(n)$ under $\Gamma_j$ can be taken to be the same as that of the representation $\F$ of $\so(8)$.\\
The schematic form of the graph near nodes $i$ and $j$ is displayed in (\ref{sf}) where $m=8$.
\item Consider a situation, where we have two nodes $i$ and $j$ such that $\Omega^{ii}=1$, $\Omega^{jj}=2$, $\Omega^{ij}=-1$, $\fg_i=\sp(n)$ and $\fg_j=\su(2n+8)$. The matter content between $\sp(n)$ and $\su(2n+8)$ is a hyper in bifundamental. In this case, the $6d$ $\sp(n)$ gauge algebra requires the input of a discrete theta angle$\theta$  which takes values $0,\pi$. For $\theta=\pi$, we need to account for the charge of BPS instanton string for $\fg_i=\sp(n)$ under its own center $\Gamma_i=\Z_2$, and the charge is $1$.\\
The graph near the nodes $i$ and $j$ takes the following schematic form
\be
\begin{tikzpicture}
\node (v1) at (-0.5,0.5) {$2$};
\node (v4) at (-0.5,1) {$\su(2n+8)$};
\begin{scope}[shift={(2.8,0.05)}]
\node at (-0.5,0.9) {$\fg_k$};
\node (v2) at (-0.5,0.45) {$\Omega^{kk}$};
\end{scope}
\node (v3) at (0.9,0.5) {\tiny{$-\Omega^{jk}$}};
\draw (v1)--(v3);
\draw (v2)--(v3);
\begin{scope}[shift={(-2.5,0.05)}]
\node (v2_1) at (-0.5,0.45) {$1$};
\end{scope}
\begin{scope}[shift={(0,1.95)}]
\node (v2_2) at (-0.5,0.45) {$\Omega^{ll}$};
\node at (-0.5,0.9) {$\fg_l$};
\end{scope}
\draw  (v2_2) edge (v4);
\draw  (v2_1) edge (v1);
\node at (-3,1) {$\sp(n)_\pi$};
\end{tikzpicture}
\ee
where we have displayed the theta angle for $\sp(n)$ which is relevant since all the $2n+8$ fundamental hypers of $\sp(n)$ are gauged by an $\su$ gauge algebra.
\een
The fact that BPS strings carry non-trivial charges under $\fg_i$ (and hence $\Gamma_i$) in the first three of the above four cases is a known fact in the literature. On the other hand, the fact that the above four cases are the \emph{only cases} where one needs to account for the charges of BPS strings under $\Gamma_i$ requires a justification, which we will provide in Section \ref{5KG}.

In any case, let us address a few pressing questions that might arise upon a reading of the above list:
\ben
\item First, it is possible, in the context of $6d$ SCFTs and LSTs, to have two nodes $i$ and $j$ with $\Omega^{ii}=1$, $\fg_i$ trivial, $\Omega^{ij}<-1$ and $\fg_j$ non-trivial. In this case, the BPS string associated to $i$ will be charged under $\fg_j$, so why is this possibility not accounted in the above list? It turns out that in this case, the charge of the BPS string under $\Gamma_j$ is trivial. To see this, notice that the only theory where this situation occurs is the following $6d$ LST
\be
\begin{tikzpicture}
\node (v1) at (-0.5,0.5) {$4$};
\node (v4) at (-0.5,1) {$\so(8)$};
\begin{scope}[shift={(2.2,0.05)}]
\node (v2) at (-0.5,0.45) {1};
\end{scope}
\node (v3) at (0.6,0.5) {\tiny{2}};
\draw (v1)--(v3);
\draw (v2)--(v3);
\end{tikzpicture} \,,
\ee
for which $\so(8)$ is embedded into $\fe_8$ with embedding index 2. Thus, the BPS string corresponding to the right node is charged as
\be
(\S\otimes\S)\oplus(\C\otimes\C)
\ee
under $\so(8)$ which has trivial charge under the $\Z_2\times\Z_2$ center of $\so(8)$.
\item Second, how about the cases, where we have a node $i$ with $\Omega^{ii}=2$ and $\fg_i$ trivial? In this case, the set of nodes $j$ such that $\Omega^{ij}<0$ and $\fg_j$ non-trivial is either trivial, or includes a single node (which we label by $j$) with $\Omega^{ij}=-1$ and $\fg_j=\su(2)$. Moreover, the $\su(2)$ gauge algebra on node $j$ must carry a positive number of full hypers in fundamental of $\su(2)$, out of which one half-hyper must be trapped by the node $i$, i.e. the half-hyper cannot be gauged by some other gauge algebra $\fg_k$. This half-hyper completely destroys the center of $\su(2)$, and hence one does not need to account for the contribution from BPS string associated to node $i$.
\item Third, in the above list the only possibilities that arise have a half-hyper charged in a mixed representation of \emph{two} simple gauge algebras. What about the possibility of having a half-hyper charged in a mixed representation of \emph{more than two} simple gauge algebras? In the context of $6d$ SCFTs and LSTs, this possibility is only realized in the $6d$ LST with the associated $6d$ gauge theory carrying $\su(2)^3$ gauge algebra along with a half-hyper in trifundamental plus two extra full hypers in fundamental representation of each $\su(2)$. In this case, the extra full hypers break the center of each of the three $\su(2)$s and hence one does not need to consider the contributions of BPS strings.
\item Fourth, how about the cases where we have a half-hyper charged under a \emph{single} gauge algebra only? In all of these cases, it turns out that there is no $6d$ SCFT or LST where the hypermultiplet content does not already capture the contribution of BPS strings. For example, consider a node $i$ with $\Omega^{ii}=3$ and $\fg_i=\so(12)$. Any $6d$ theory containing this node contains a half-hyper charged as $\S$ of $\so(12)$ and 5 hypers charged as $\F$. Since the half-hyper in $\S$ cannot be gauged by any other gauge algebra $\fg_j$ for a $6d$ SCFT or LST, the $\Z_2^2$ center of $\so(12)$ is broken down to the $\Z_2$ subgroup under which $\F$ and $\C$ reps of $\so(12)$ have charge 1. It turns out that there is no way to gauge the 5 hypers in $\F$ and to simultaneously complete the node $i$ into a $6d$ SCFT or LST such that the above $\Z_2$ subgroup of the center of $\so(12)$ would survive as 1-form symmetry. Thus, the center of $\so(12)$ is already completely broken by the hypermultiplet content, and we do not get to the point where we need to discuss the charge of BPS string associated to $i$ under $\Gamma_i$.
\een

\subsubsection{Examples}\label{1ex}
\ni\ubf{Example 1}: Consider the $6d$ SCFT
\be
\begin{tikzpicture}
\node (v1) at (-0.5,0.5) {$4$};
\node (v4) at (-0.5,1) {$\so(4n)$};
\end{tikzpicture}
\ee
where $n\ge2$. The center of $\so(4n)$ is $\Z_2\times\Z_2$ under which fundamental, spinor and cospinor representations have charges $(1,1)$, $(1,0)$ and $(0,1)$ respectively. The above $6d$ SCFT contains $4n-8$ hypers in fundamental representation. For $n=2$, there are no hypers and we find
\be
\cO=\Z_2\times\Z_2 \,.
\ee
For $n>2$, the fundamental hypers are uncharged under only a diagonal combination of the two $\Z_2$s and thus
\be
\cO=\Z_2 \,.
\ee

For the $6d$ SCFT
\be
\begin{tikzpicture}
\node (v1) at (-0.5,0.5) {$4$};
\node (v4) at (-0.5,1) {$\so(4n+2)$};
\end{tikzpicture}
\ee
the center is $\Z_4$ under which fundamental has charge $2$ and spinor/cospinor have charges $\pm1$. The $6d$ SCFT contains $4n-6$ hypers and $n\ge2$ for the theory to exist. The presence of fundamental hypers implies that the 1-form symmetry for this theory is
\be
\cO=\Z_2 \,.
\ee

In all of the cases considered in this example, there is no extra breaking induced by the instanton string.

\vspace{8pt}

\ni\ubf{Example 2}: Consider the $6d$ SCFT
\be
\begin{tikzpicture}
\node (v1) at (-0.5,0.5) {$4$};
\node (v4) at (-0.5,1) {$\so(2n)$};
\begin{scope}[shift={(2.2,0.05)}]
\node (v2) at (-0.5,0.45) {1};
\end{scope}
\node (v4) at (1.7,1) {$\sp(2n-8)$};
\draw  (v1) edge (v2);
\end{tikzpicture}
\ee
Consider first the $n>4$ case, for which we have a half-hyper in the bifundamental and $3n-8$ full hypers in fundamental of $\sp(2n-8)$. The presence of fundamentals of $\sp(2n-8)$ breaks the $\Z_2$ center 1-form symmetry associated to $\sp(2n-8)$ down to $\Z_1$. And the presence of bifundamental breaks the center 1-form symmetry associated to $\so(2n)$ down to the $\Z_2$ subgroup under which fundamental representation is uncharged.

However, this is not the end of story, as the BPS instanton string associated to the $\sp(2n-8)$ has non-trivial charge under the above $\Z_2$ subgroup of the center of $\so(2n)$. Thus, we find that the 1-form symmetry for the above $6d$ SCFT is trivial. That is,
\be
\cO=\Z_1 \,.
\ee

For $n=4$, $\sp(2n-8)=\sp(0)$ denotes that there is no gauge algebra associated to the right node and we can write the quiver as
\be
\begin{tikzpicture}
\node (v1) at (-0.5,0.5) {$4$};
\node (v4) at (-0.5,1) {$\so(8)$};
\begin{scope}[shift={(2,0.05)}]
\node (v2) at (-0.5,0.45) {1};
\end{scope}
\draw  (v1) edge (v2);
\end{tikzpicture}
\ee
The potential 1-form symmetry is $\Z_2\times\Z_2$ coming from the center of $\so(8)$. There are no hypermultiplets, but we again have to account for the BPS string associated to the right node. This string is charged as the adjoint of the total flavor symmetry $\fe_8$ associated to the right node. The $\so(8)$ gauge algebra embeds into $\fe_8$ such that the adjoint of $\fe_8$ decomposes into a representation of $\so(8)$ which contains both the spinor and cospinor representations. Thus, both the $\Z_2$s are broken by this BPS string and we again obtain
\be
\cO=\Z_1 \,.
\ee

Consider also the $6d$ LST
\be
\begin{tikzpicture}
\node (v1) at (-0.5,0.5) {$4$};
\node (v4) at (-0.5,1) {$\so(2n)$};
\begin{scope}[shift={(2.2,0.05)}]
\node (v2) at (-0.5,0.45) {1};
\end{scope}
\node (v3) at (0.6,0.5) {\tiny{2}};
\draw (v1)--(v3);
\draw (v2)--(v3);
\node at (1.7,1) {$\sp(n-8)$};
\end{tikzpicture}
\ee
whose matter content is a full hyper rather than a half-hyper in the bifundamental of the two algebras. According to our general discussion above, due to the presence of a full hyper, we don't need to consider the contribution of BPS instanton strings. Any element of the center $\Gamma_{\so}$ of $\so(2n)$ that acts non-trivially on the representation $\F$ of $\so(2n)$ can be combined with the generator of the center $\Z_2$ of $\sp(n-8)$ to produce an element of the 1-form symmetry group of the above theory. Thus, we find that
\be
\cO\simeq\Gamma_{\so} \,.
\ee

\vspace{8pt}

\ni\ubf{Example 3}: Consider the $6d$ SCFT
\be
\begin{tikzpicture}
\node (v1) at (-0.5,0.5) {$4$};
\node (v4) at (-0.5,1) {$\so(2n)$};
\begin{scope}[shift={(2.2,0.05)}]
\node (v2) at (-0.5,0.45) {2};
\end{scope}
\node (v3) at (0.6,0.5) {\tiny{2}};
\draw (v1)--(v3);
\draw (v2)--(v3);
\node (v4) at (1.7,1) {$\su(2n-8)$};
\end{tikzpicture} \,
\ee
where $n\ge8$. The theory contains a bifundamental hyper plus $2n-16$ fundamental hypers for $\su(2n-8)$. Let us first consider the case $n>8$. Then, the $2n-16$ fundamental hypers of $\su(2n-8)$ completely destroy the center $\Z_{2n-8}$ 1-form symmetry associated to $\su(2n-8)$. As above, the bifundamental hyper leaves only a $\Z_2$ 1-form symmetry out of the center 1-form symmetry associated to $\so(2n)$. The BPS strings do not contribute to any additional breaking of the potential 1-form symmetry since the theory does not contain any half-hypers in mixed representation of $\so(2n)\oplus\su(2n-8)$. Thus, the 1-form symmetry is
\be
\cO=\Z_2 \,,
\ee
for $n>8$.

Now consider the case $n=8$. We can combine the order two element in the center $\Z_8$ associated to $\su(2n-8)=\su(8)$ with the generators of the two $\Z_2$s in the center of $\so(2n)=\so(16)$ to obtain two $\Z_2$ symmetries under which the hypermultiplet content is uncharged. Due to the same reason as for the case $n>8$, the BPS strings do not further reduce the 1-form symmetry in the $n=8$ case as well. Thus, the above $6d$ SCFT for $n=8$ has 1-form symmetry
\be
\cO=\Z_2\times\Z_2 \,.
\ee

\vspace{8pt}

\ni\ubf{Example 4}: Consider the $6d$ SCFT
\be
\begin{tikzpicture}
\node (v1) at (-0.5,0.5) {$4$};
\node (v4) at (-0.5,1) {$\so(2n+8)$};
\begin{scope}[shift={(2,0.05)}]
\node (v2) at (-0.5,0.45) {1};
\end{scope}
\draw  (v1) edge (v2);
\node (v1_1) at (3.5,0.5) {$4$};
\draw  (v2) edge (v1_1);
\node (v4_1) at (3.5,1) {$\so(2n+8)$};
\node (v4_1) at (1.5,1) {$\sp(n)$};
\end{tikzpicture}
\ee
which makes sense for $n\ge0$. Consider first the case of $n>0$. Then, the hypermultiplet content of the theory is
\be
\half(\F,\F,1)\oplus\half(1,\F,\F)\oplus n(\F,1,1)\oplus n(1,1,\F) \,,
\ee
where $\F$ denotes the fundamental representation. This breaks the $\Z_2$ center of $\sp(n)$, but leaves a $\Z_2$ element inside the center of each $\so(2n+8)$ unbroken. The unbroken $\Z_2$ inside $\so(2n+8)$ acts non-trivially on the spinor and cospinor representations but acts trivially on the fundamental representation. The BPS instanton string for $\sp(n)$ has charge $(1,1)$ under the unbroken $\Z_2^2$ potential 1-form symmetry coming from the two $\so(2n+8)$ gauge algebras. Thus we see that only a diagonal combination of the two surviving $\Z_2$s associated to the two $\so(2n+8)$s survives. That is, the 1-form symmetry for $n>0$ is
\be
\cO=\Z_2 \,.
\ee
Notice that if one of the two $\so(2n+8)$ was not gauged, then we would have obtained a trivial 1-form symmetry as discussed in an example above.

Now consider the case of $n=0$ for which we can write the quiver as
\be
\begin{tikzpicture}
\node (v1) at (-0.5,0.5) {$4$};
\node (v4) at (-0.5,1) {$\so(8)$};
\begin{scope}[shift={(1.5,0.05)}]
\node (v2) at (-0.5,0.45) {1};
\end{scope}
\draw  (v1) edge (v2);
\node (v1_1) at (2.5,0.5) {$4$};
\draw  (v2) edge (v1_1);
\node (v4_1) at (2.5,1) {$\so(8)$};
\end{tikzpicture}
\ee
This theory contains no charged hypermultiplets. But the BPS string associated to the middle node is charged under the adjoint of its $\fe_8$ flavor symmetry, which decomposes under the two $\so(8)$s as
\be
(\A,1)\oplus(1,\A)\oplus(\F,\F)\oplus(\S,\S)\oplus(\C,\C) \,
\ee
where $\A$ denotes the adjoint representation. Thus we see that the BPS string is left invariant by a diagonal combination of the centers of the two $\so(8)$. Thus, the 1-form symmetry is
\be
\cO=\Z_2\times\Z_2 \,.
\ee

This result can be extended to the $6d$ SCFT
\be
\begin{tikzpicture}
\node (v1) at (-0.5,0.5) {$4$};
\node (v4) at (-0.5,1) {$\so(8)$};
\begin{scope}[shift={(1.5,0.05)}]
\node (v2) at (-0.5,0.45) {1};
\end{scope}
\draw  (v1) edge (v2);
\node (v1_1) at (2.5,0.5) {$4$};
\draw  (v2) edge (v1_1);
\node (v4_1) at (2.5,1) {$\so(8)$};
\begin{scope}[shift={(4.5,0.05)}]
\node (v2_1) at (-0.5,0.45) {1};
\end{scope}
\node (v1_1_1) at (7,0.5) {$4$};
\node (v4_1_1) at (7,1) {$\so(8)$};
\node (v1_1_2) at (5.5,0.5) {$\cdots$};
\draw  (v1_1) edge (v2_1);
\draw  (v2_1) edge (v1_1_2);
\draw  (v1_1_2) edge (v1_1_1);
\end{tikzpicture}
\ee
for which only a diagonal combination of the centers of all the $\so(8)$s survives, thus leading to
\be
\cO=\Z_2\times\Z_2 \,.
\ee

\vspace{8pt}

\ni\ubf{Example 5}: Consider the $6d$ SCFT
\be
\begin{tikzpicture}
\node (v1) at (-0.5,0.5) {$6$};
\node (v4) at (-0.5,1) {$\fe_6$};
\begin{scope}[shift={(1.5,0.05)}]
\node (v2) at (-0.5,0.45) {1};
\end{scope}
\draw  (v1) edge (v2);
\node (v1_1) at (2.5,0.5) {$3$};
\draw  (v2) edge (v1_1);
\node (v4_1) at (2.5,1) {$\su(3)$};
\end{tikzpicture}
\ee
which carries no charged hypers and for which the BPS string associated to the middle node is charged under $\fe_6\oplus\su(3)$ as
\be
(\A,1)\oplus(1,\A)\oplus(\F,\bar\F)\oplus(\bar\F,\F) \,,
\ee
where $\F=\mathbf{27}$ for $\fe_6$. This is left invariant by a diagonal $\Z_3$ combination of the $\Z_3$ centers associated to $\fe_6$ and $\su(3)$, thus leading to the final result
\be
\cO=\Z_3
\ee

This result can be extended to the $6d$ SCFTs
\be
\begin{tikzpicture}
\node (v1) at (-0.5,0.5) {$3$};
\node (v4) at (-0.5,1) {$\su(3)$};
\begin{scope}[shift={(1.5,0.05)}]
\node (v2) at (-0.5,0.45) {1};
\end{scope}
\draw  (v1) edge (v2);
\node (v1_1) at (2.5,0.5) {$6$};
\draw  (v2) edge (v1_1);
\node (v4_1) at (2.5,1) {$\fe_6$};
\begin{scope}[shift={(4.5,0.05)}]
\node (v2_1) at (-0.5,0.45) {1};
\end{scope}
\node (v1_1_1) at (7,0.5) {$3$};
\node (v4_1_1) at (7,1) {$\su(3)$};
\node (v1_1_2) at (5.5,0.5) {$\cdots$};
\draw  (v1_1) edge (v2_1);
\draw  (v2_1) edge (v1_1_2);
\draw  (v1_1_2) edge (v1_1_1);
\begin{scope}[shift={(-1.5,0.05)}]
\node (v2_2) at (-0.5,0.45) {1};
\end{scope}
\node (v1_1_3) at (-3.5,0.5) {$6$};
\node (v4_1_2) at (-3.5,1) {$\fe_6$};
\draw  (v1_1_3) edge (v2_2);
\draw  (v2_2) edge (v1);
\end{tikzpicture}
\ee
and
\be
\begin{tikzpicture}
\node (v1) at (-0.5,0.5) {$3$};
\node (v4) at (-0.5,1) {$\su(3)$};
\begin{scope}[shift={(1.5,0.05)}]
\node (v2) at (-0.5,0.45) {1};
\end{scope}
\draw  (v1) edge (v2);
\node (v1_1) at (2.5,0.5) {$6$};
\draw  (v2) edge (v1_1);
\node (v4_1) at (2.5,1) {$\fe_6$};
\begin{scope}[shift={(4.5,0.05)}]
\node (v2_1) at (-0.5,0.45) {1};
\end{scope}
\node (v1_1_1) at (7,0.5) {$6$};
\node (v4_1_1) at (7,1) {$\fe_6$};
\node (v1_1_2) at (5.5,0.5) {$\cdots$};
\draw  (v1_1) edge (v2_1);
\draw  (v2_1) edge (v1_1_2);
\draw  (v1_1_2) edge (v1_1_1);
\begin{scope}[shift={(-1.5,0.05)}]
\node (v2_2) at (-0.5,0.45) {1};
\end{scope}
\node (v1_1_3) at (-3.5,0.5) {$6$};
\node (v4_1_2) at (-3.5,1) {$\fe_6$};
\draw  (v1_1_3) edge (v2_2);
\draw  (v2_2) edge (v1);
\end{tikzpicture}
\ee
for which again only a diagonal $\Z_3$ combination of all the centers survives, leading to
\be
\cO=\Z_3
\ee

\vspace{8pt}

\ni\ubf{Example 6}: Consider the following LST arising in the frozen phase of F-theory
\be
\begin{tikzpicture}
\node (v1) at (-1,0.5) {$1$};
\node (v4) at (-1,1) {$\sp(n)_\pi$};
\begin{scope}[shift={(2,0.05)}]
\node (v2) at (-0.5,0.45) {2};
\end{scope}
\draw  (v1) edge (v2);
\node (v1_1) at (4.5,0.5) {$4$};
\node (v4_1) at (4.5,1) {$\so(2n+16)$};
\node (v4) at (1.5,1) {$\su(2n+8)$};
\node (v3) at (3,0.5) {\tiny{2}};
\draw  (v2) edge (v3);
\draw  (v3) edge (v1_1);
\end{tikzpicture} \,,
\ee
for $n>0$, where the theta angle for $\sp(n)$ is relevant since all of the $2n+8$ fundamental hypers of $\sp(n)$ have been gauged by $\su(2n+8)$ gauge algebra, and we have chosen this theta angle to be $\pi$. The hypermultiplet content forms a representation
\be
(\F,\F,1)\oplus(1,\F,\F) \,,
\ee
of $\sp(n)\oplus\su(2n+8)\oplus\so(2n+16)$. The potential center 1-form symmetry is $\Gamma:=\Z_2\times\Z_{2n+8}\times\Gamma_{\so}$ where $\Z_2$ factor is the center of $\sp(n)$, $\Z_{2n+8}$ factor is the center of $\su(2n+8)$ and $\Gamma_{\so}$ is the center of $\so(2n+16)$, where $\Gamma_{\so}=\Z_4$ if $n$ is odd and $\Gamma_{\so}=\Z_2^2$ when $n$ is even). This potential 1-form symmetry is broken by the above hyper content to a subgroup $\tilde\Gamma$ of $\Gamma$. It turns out that $\tilde\Gamma$ is isomorphic to $\Gamma_{\so}$ with the generators of $\tilde\Gamma$ being obtained by combining the generators of the $\Gamma_{\so}$ factor of $\Gamma$ combined with the order 2 element in the $\Z_{2n+8}$ factor of $\Gamma$ combined with the generator of $\Z_2$ factor of $\Gamma$.

However, the BPS string associated to the $\sp(n)$ node has charge 1 under the $\Z_2$ factor of $\Gamma$ since the theta angle for $\sp(n)$ is $\pi$, and hence the $\tilde\Gamma$ potential 1-form symmetry is completely broken since all the generators of $\tilde\Gamma$ involve the generator of the $\Z_2$ factor of $\Gamma$. We find that the above LST has
\be
\cO=\Z_1 \,.
\ee

\section{1-form symmetry of $5d$ $\cN=1$ theories}\label{5}
In this section, our aim is to study higher-form symmetries of $5d$ $\cN=1$ theories. More precisely, we aim to study mass-deformations of $5d$ SCFTs and circle compactifications of $6d$ SCFTs and LSTs. 

Just as in the previous section, we would like to argue that it is sufficient for us to focus on a class of $5d$ theories, which admit only one kind of higher-form symmetries, namely 1-form symmetries \footnote{Just like the case of 1-form and 2-form symmetries of $6d$ theories, these 1-form symmetries of $5d$ theories will also be spontaneously broken in all kinds of vacua we discuss below.}. The argument is again that all known $5d$ theories arise by discrete gaugings of the above class of theories\footnote{See however \cite{Closset:2020scj} for some proposed counter-examples. In these cases, there are 3-form symmetries, whose interpretation remains to be fully understood in terms of the classification of 5d SCFTs. }. Moreover, all the known $5d$ theories in the above class admit a geometric construction in M-theory which we will be using to study these theories. The geometric constructions that we will consider require extra discrete data that we fix by demanding that all the non-compact complex curves can be wrapped by M2-branes. This severely limits the non-compact complex surfaces that can be wrapped by M5-branes. See \cite{Morrison:2020ool, Albertini:2020mdx} for more discussion about this discrete data. It is this above mentioned choice of discrete data that gives rise to the $5d$ theories in the above mentioned class of $5d$ theories that we will be studying.

\subsection{1-form symmetry from the Coulomb branch}
At a generic point on its Coulomb branch, a $5d$ $\cN=1$ theory flows to a $5d$ $\cN=1$ abelian gauge theory with gauge group $U(1)^r$, where $r$ is often called as the rank of the original $5d$ $\cN=1$ theory. We can choose a basis for $U(1)^r$ such that the $U(1)^r$ charges of the line defects and dynamical particles in the theory lie in a lattice generated by primitive Wilson lines $W_i$ having charge $+1$ under $U(1)_i$ gauge group and charge $0$ under $U(1)_j$ gauge group for $j\neq i$.

Each $U(1)_i$ gauge group gives rise to a potential $U(1)$ 1-form symmetry, and we can identify the actual 1-form symmetry group $\cO$ of the $5d$ $\cN=1$ theory as the elements of these potential $U(1)$ 1-form symmetries under which all the BPS (and massless) particles are uncharged.

\subsubsection{1-form symmetry from M-theory geometry}\label{5G}
The above discussed procedure of determining the 1-form symmetry of a $5d$ $\cN=1$ theory from its Coulomb branch is easy to implement if the $5d$ $\cN=1$ theory admits a geometric construction in M-theory. In such a construction, the Coulomb branch of $5d$ $\cN=1$ theory is constructed by compactifying M-theory on a non-compact Calabi-Yau threefold (CY3). 

The CY3 contains a collection of irreducible compact Kahler surfaces $S_i$. Decomposing the M-theory 3-form gauge field in terms of a basis of 2-forms associated to $S_i$ leads to a collection of 1-forms $A_i$ which are identified as the gauge fields for gauge groups $U(1)_i$. The CY3 also contains compact holomorphic curves which lead to dynamical BPS particles via compactification of M2-branes on these curves. The charge of a particle arising from a curve $C$ under $U(1)_i$ is given by the intersection number $C\cdot S_i$.

Typically, the surfaces $S_i$ can be identified as blowups of Hirzebruch surfaces or blowups of $\P^2$. Moreover, the CY3 can often be presented in a form such that each curve $C$ can be written as a linear combination of compact curves living inside $S_i$. The intersection number $C\cdot S_i$ can then be traced to intersection theory of Hirzebruch surfaces and $\P^2$.

To do this, let $\alpha$ parametrize different intersections between $S_i$ and $S_j$ for $i\neq j$. Then the locus of $\alpha^{\text{th}}$ intersection can be identified as a compact curve $C_{ij}^{(\alpha)}$ living in $S_i$ and a compact curve $C_{ji}^{(\alpha)}$ living in $S_j$. In other words, we say that the $\alpha^{\text{th}}$ intersection between $S_i$ and $S_j$ is produced by identifying the curve $C_{ij}^{(\alpha)}$ living in $S_i$ with the curve $C_{ji}^{(\alpha)}$ living in $S_j$. We refer to $C_{ij}^{(\alpha)}$ and $C_{ji}^{(\alpha)}$ as the gluing curves corresponding to this intersection. Moreover, let us define the \emph{total} gluing curves for the intersections of $S_i$ and $S_j$ as $C_{ij}:=\sum_\alpha C_{ij}^{(\alpha)}$ and $C_{ji}:=\sum_\alpha C_{ji}^{(\alpha)}$.

Similarly, different self-intersections of a surface $S_i$ can be obtained by gluing $C_i^{(\alpha)}$ with $D_i^{(\alpha)}$ where $C_i^{(\alpha)}$ and $D_i^{(\alpha)}$ are curves living in $S_i$. In this case, we identify the total \emph{self-gluing} curve as $C_i:=\sum_\alpha C_i^{(\alpha)}+\sum_\alpha D_i^{(\alpha)}$.

If a compact curve $C$ lives in $S_i$ then its intersection number with $S_j$ for $j\neq i$ can be written as
\be
C\cdot S_j=(C\cdot C_{ij})_{S_i} \,,
\ee
where the brackets with a subscript $S_i$ represents the fact that the intersection can be taken inside the surface $S_i$ without regard for the details of the rest of the CY3. On the other hand, the intersection number of $C$ with $S_i$ can be written as
\be
C\cdot S_i=(C\cdot K_i)_{S_i}+(C\cdot C_i)_{S_i}=2g(C)-2-(C\cdot C)_{S_i}+(C\cdot C_i)_{S_i}
\ee
where $K_i$ is the canonical divisor of $S_i$ and we have used the adjunction formula (applied to the surface $S_i$) to write its intersection with $C$ in terms of the self-intersection of $C$ (inside $S_i$) and the genus $g(C)$ of $C$.

The upshot of the above discussion is that we can reduce the calculation of $U(1)_i$ charges of various dynamical particles in the $5d$ $\cN=1$ theory to the calculation of some intersection numbers inside the surfaces $S_i$, where an intersection number \emph{inside} $S_i$ can be computed without regard for the details of the rest of the CY3. Now we only need to discuss the intersection theory of curves inside a fixed surface $S_i$.

As we remarked above, each $S_i$ is either a blowup of a Hirzebruch surface or a blowup of $\P^2$. The first homology of a blowup of Hirzebruch surface can be described in terms of curves $e$, $f$ and $x_i$, where $e$ is the homology class of the total transform (under all blowups) of the base $\P^1$ of the Hirzebruch surface, $f$ is the homology class of the total transform (under all blowups) of a fiber $\P^1$ of the Hirzebruch surface, and $x_i$ is the homology class of the total transform (under subsequent\footnote{For our convenience, when we consider concrete geometries below, we will \emph{not} adopt the order that the blowup $j$ is performed after blowup $i$ if $j>i$.} blowups $j>i$) of the exceptional $\P^1$ introduced by the $i^{\text{th}}$ blowup.

Similarly, the first homology of a blowup of $\P^2$ can be described in terms of curves $l$ and $x_i$, where $l$ is the homology class of the total transform (under all blowups) of a $\P^1$ inside $\P^2$, and $x_i$ is the homology class of the total transform (under subsequent blowups $j>i$) of the exceptional $\P^1$ introduced by the $i^{\text{th}}$ blowup.

The intersection numbers between these curves in the case of a Hirzebruch surface $\bF_n$ of degree $n$ are
\begin{align}
e\cdot e &= -n\\
f\cdot f &= 0\\
x_i\cdot x_j &= -\delta_{ij}\\
e\cdot f &= +1\\
x_i\cdot e &= 0\\
x_i\cdot f &= 0 \,.
\end{align}
We will also use the $h$ curve which is defined as
\be
h:=e+nf \,.
\ee
On the other hand, the intersection numbers in the case of $\P^2$ are
\begin{align}
l\cdot l &= +1\\
x_i\cdot x_j &= -\delta_{ij}\\
x_i\cdot l &= 0 \,.
\end{align}
Using the above information, we can determine the $U(1)_i$ charges of any dynamical particle on the Coulomb branch of the $5d$ $\cN=1$ theory $\fT$ in consideration. Similar to the case in Section \ref{6T}, the 1-form symmetry group $\cO$ for $\fT$ can be computed from the point of view of its Pontryagin dual. For this purpose, let $\Z^r$ be the lattice of possible $U(1)_i$ charges. Then, let $\cC$ be a set of curves defined as follows:

For each $S_i$, which is a blowup of a Hirzebruch surface, we add the curves $e,f,x_i$ into $\cC$, and for each $S_i$, which is a blowup of $\P^2$, we add the curves $l,x_i$ into $\cC$.\\
Let $\alpha$ parametrize different elements of $\cC$. Then, the $U(1)_i$ charges of elements of $\cC$ define the \emph{charge matrix} $Q^{\alpha i}$, which can be used to describe $\cO$ as the Pontryagin dual of the quotient lattice\footnote{This result was first derived in \cite{Morrison:2020ool}.}
\be
\frac{\Z^r}{[Q^{\alpha i}]\cdot \Z^{r}}=\bigoplus_{i=1}^r~\frac{\Z}{n_i\Z} \,,
\ee
where $n_i:=\tilde Q^{ii}$ and $\tilde Q^{\alpha i}$ is the Smith normal form of $Q^{\alpha i}$.

If the $5d$ $\cN=1$ theory is a $5d$ SCFT or a compactification of a $6d$ SCFT (twisted or untwisted) on a circle of finite non-zero radius, then each $n_i>0$, and we can write the Pontryagin dual as
\be
\cO=\prod_{i=1}^r~\Z_{n_i} \,,
\ee
with $\Z_1$ being the trivial group.

\subsection{1-form symmetry of $5d$ $\cN=1$ non-abelian gauge theories}\label{5N}
As in Section \ref{6O}, the 1-form symmetry of a non-abelian $5d$ $\cN=1$ gauge theory with gauge algebra $\fg=\oplus_i \fg_i$ (where $\fg_i$ are simple) can be described as a subgroup $\cO$ of $\prod_i\Gamma_i$ where $\Gamma_i$ is the center of $\fg_i$. One necessary condition on $\cO$ is that its elements should leave all the (full or half) hypermultiplets invariant. As in Section \ref{6O}, we also need to include the instantonic excitations. In that section, the effect of these excitations was captured by requiring that the fundamental BPS instanton strings be uncharged under elements of $\cO$. In the case of $5d$ $\cN=1$ theories, the effect of instantonic excitations is captured by requiring that BPS instanton particles are left invariant by elements of $\cO$.

Some examples of instantonic contributions to (the breaking of) 1-form symmetry in $5d$ theories were already studied in \cite{Morrison:2020ool}. Two such examples are obtained by considering a pure $5d$ $\cN=1$ gauge theory with a \emph{simple} gauge algebra $\fg=\su(n),\sp(n)$. As discussed in the above reference, for a pure $\su(n)$ theory with Chern-Simons (CS) level $k$, the instantonic contributions are captured by accounting for an instanton particle of charge $k~(\text{mod}~n)$ under the center $\Z_n$ of $\su(n)$; and for a pure $\sp(n)$ theory with theta angle $\theta=m\pi~(\text{mod}~2\pi)$, the instantonic contributions are captured by accounting for an instanton particle of charge $m~(\text{mod}~2)$ under the center $\Z_2$ of $\sp(n)$.

In this subsection, we will discuss other examples where instantonic contributions are relevant to the discussion of 1-form symmetry of $5d$ gauge theories. To this end, we will employ the M-theory construction of these $5d$ gauge theories.

\subsubsection{1-form symmetry of non-abelian gauge theories from geometry}\label{5GG}
In Section \ref{5G}, we discussed geometric constructions of Coulomb branches of $5d$ $\cN=1$ theories. At special loci in the Coulomb branch, the low-energy theory enhances from an abelian gauge theory to a non-abelian gauge theory such that in the vicinity of such a locus we can regard the abelian gauge theory as arising on the Coulomb branch of the non-abelian gauge theory.

Let us consider a locus where a non-abelian gauge theory with a semi-simple gauge algebra $\fg$ arises. In the vicinity of this locus, the M-theory geometry can be represented in the following special form (see \cite{Bhardwaj:2019ngx,Bhardwaj:2020gyu} for more details):\\
We can represent each surface $S_i$ as a blowup of a Hirzebruch surface such that the \emph{intersection matrix} $M_{ij}$ defined by
\be\label{Cartan}
M_{ij}:=-f_i\cdot S_j \,,
\ee
(where $f_i$ denotes (the homology class of) a fiber $\P^1$ of Hirzebruch surface $S_i$) can be identified as the Cartan matrix of $\fg$.\\
The hypermultiplet content of the non-abelian gauge theory is encoded in the blowups and gluing curves. The details of this encoding can be found in \cite{Bhardwaj:2019ngx,Bhardwaj:2020gyu}. Here we will only need to consider special cases of the general case analyzed there.

(\ref{Cartan}) establishes a one-to-one correspondence between the nodes in the Dynkin diagram of $\fg$ and the surfaces $S_i$. Let the semi-simple gauge algebra $\fg$ decompose into simple factors as $\fg=\oplus_\mu\fg_\mu$. Let $S_i^\mu$ be the surfaces corresponding to $\fg_i^\mu$.

(\ref{Cartan}) implies that the total gluing curve $C_{ij}$ for $i\neq j$ can be written as
\be
C_{ij}=-M_{ij}e_i+\beta_{ij}f_i+\sum_m\gamma_{ijm}x_{im}
\ee
for some undetermined coefficients $\beta_{ij}$ and $\gamma_{ijm}$ where $x_{im}$ are the blowups living in the Hirzebruch surface $S_i$. Using the above form for $C_{ij}$ and structure of Cartan matrix $M_{ij}$, we can find a (non-unique) surface $\tilde S^\mu$ among the surfaces $S^i_\mu$ such that we can write
\be\label{inst}
e_i^\mu\sim n_i^\mu\tilde e^\mu+\cdots \,,
\ee
where the $\sim$ sign denotes the curves on the two sides are same inside the homology of the full threefold; $\tilde e^\mu$ is the $e$ curve for the surface $\tilde S^\mu$; $n_i^\mu$ are strictly positive integers; and the omitted terms denoted by dots include contribution only from fibers and blowups living inside surfaces $S_i^\mu$ for various $i$. An explicit choice for $\tilde e_\mu$ for various simple Lie algebras will be provided later in this subsection. This result (\ref{inst}) will be very helpful for us in determining the contribution of instantons to the 1-form symmetry, but let us keep it aside for some time and turn to the discussion of the realization of center symmetry in terms of surfaces $S_i$.

For each $\mu$ we have surfaces $S_i^\mu$ for $i=1,\cdots,r_\mu$ where $r_\mu$ is the rank of $\fg_\mu$. Consider the lattice $\Lambda_S^\mu\simeq\Z^{r_\mu}$ spanned by $S_i^\mu$ and the lattice $\Lambda_f^\mu\simeq\Z^{r_\mu}$ spanned by $f_i^\mu$. We claim that we can change basis inside $\Lambda_S^\mu$ from $S_i^\mu$ to $S_a^\mu$ (which are some linear combinations of $S_i^\mu$) with $a=1,\cdots,r_\mu$, and the basis inside $\Lambda_f^\mu$ from $f_i^\mu$ to $f_a^\mu$ (which are some linear combinations of $f_i^\mu$) with $a=1,\cdots,r_\mu$, such that
\begin{align}
-f_a^\mu\cdot S_b^\mu&=\delta_{ab}\\
-f_c^\mu\cdot S_b^\mu&=0\\
-f_a^\mu\cdot S_c^\mu&=0
\end{align}
for $a,b>1$ and $c=1$ if $\fg_\mu\neq\so(4n)$; and $a,b>2$ and $c=1,2$ if $\fg_\mu=\so(4n)$ for some $n$. Furthermore,
\be
-f_1^\mu\cdot S^1_\mu=N_\mu
\ee
for $\fg_\mu\neq\so(4n)$ where $\Z_{N_\mu}$ is the center of $\fg_\mu$, and
\be
-f_a^\mu\cdot S_b^\mu=2\delta_{ab}
\ee
for $\fg_\mu=\so(4n)$ where $a,b\in\{1,2\}$. More importantly, these results imply that if $\fg_\mu\neq\so(4n)$, then
\be\label{C1}
-f_i^\mu\cdot S_{a=1}^\mu=k_i^\mu N_\mu
\ee
for some integers $k_i^\mu$ having gcd 1. Similarly, if $\fg_\mu=\so(4n)$, then
\be\label{C2}
-f_i^\mu\cdot S_{a}^\mu=2k_{ia}^\mu
\ee
for $a=1,2$ and some integers $k_{i1}^\mu$ having gcd 1 and some integers $k_{i2}^\mu$ having gcd 1.

The upshot of the above analysis is that we have changed the basis of potential 1-form symmetries from $U(1)^\mu_i$ to $U(1)^\mu_a$ such that the W-bosons $f_i^\mu$ break $U(1)^\mu_a$ down to the center $\Gamma_\mu$ of $\fg_\mu$. For $\fg_\mu\neq\so(4n)$, the \emph{center 1-form symmetry} arises from the $U(1)_{a=1}^\mu$ associated to the surface $S_{a=1}^\mu$. For $\fg_\mu=\so(4n)$, the \emph{center 1-form symmetry} has two factors which arise from the $U(1)_{a=1}^\mu$ and $U(1)_{a=2}^\mu$ associated to the surfaces $S_{a=1}^\mu$ and $S_{a=2}^\mu$. (\ref{C1}) and (\ref{C2}) simply state that the W-bosons have a charge
\be
0~(\text{mod}~n)
\ee
under $U(1)_a^\mu$ where $n$ is the order of the \emph{center symmetry} associated to $U(1)_a^\mu$.

Let us now provide an explicit identification of surfaces $S_{a=1}^\mu$ for various possible simple Lie algebras $\fg_\mu\neq\so(4n)$ and an explicit identification of surfaces $S_{a=1,2}^\mu$ for $\fg_\mu=\so(4n)$. As we have discussed above, these surfaces generate the center 1-form symmetries associated to $\fg_\mu$. We leave an explicit identification of $f_b^\mu$ and $S_a^\mu$ for other values of $a$ to the reader.
\bit
\item For $\fg_\mu=\su(n)$, label the nodes in the Dynkin diagram as
\be
\begin{tikzpicture}[scale=1.5]
\draw[fill=black]
(0,0) node (v1) {} circle [radius=.1]
(1,0) node (v3) {} circle [radius=.1]
(2,0) node (v4) {} circle [radius=.1]
(4,0) node (v5) {} circle [radius=.1];
\node (v2) at (3,0) {$\cdots$};
\draw  (v1) edge (v3);
\draw  (v3) edge (v4);
\draw  (v4) edge (v2);
\draw  (v2) edge (v5);
\node at (0,-0.3) {{$1$}};
\node at (1,-0.3) {{$2$}};
\node at (2,-0.3) {{$3$}};
\node at (4,-0.3) {{$n-1$}};
\end{tikzpicture}
\ee
Then, we can take
\be
S_{a=1}^\mu=\sum_{i=1}^{n-1}iS^\mu_i \,.
\ee
Only the fiber $f_{i=n-1}^\mu$ has a non-zero charge under the $U(1)$ generated by the above surface. This fiber has charge $n$, thus reducing the $U(1)$ generated by $S^\mu_{a=1}$ to $\Z_n$, which can be identified as the center of $\su(n)$.\\
We can choose $\tilde e^\mu=e_{i=1}^\mu$.
\item For $\fg_\mu=\so(2n+1)$, label the nodes in the Dynkin diagram as
\be
\begin{tikzpicture}[scale=1.5]
\draw[fill=black]
(0,0) node (v1) {} circle [radius=.1]
(1,0) node (v3) {} circle [radius=.1]
(2,0) node (v4) {} circle [radius=.1]
(4,0) node (v5) {} circle [radius=.1]
(5,0) node (v6) {} circle [radius=.1];
\node (v2) at (3,0) {$\cdots$};
\draw  (v1) edge (v3);
\draw  (v3) edge (v4);
\draw  (v4) edge (v2);
\draw  (v2) edge (v5);
\node at (0,-0.3) {{$1$}};
\node at (1,-0.3) {{$2$}};
\node at (2,-0.3) {{$3$}};
\node at (4,-0.3) {{$n-1$}};
\node at (5,-0.3) {{$n$}};
\begin{scope}[shift={(-1,0)}]
\draw (5.15,0.025) -- (5.825,0.025) (5.825,-0.025) -- (5.15,-0.025);
\draw (5.5,0.1) -- (5.6,0) -- (5.5,-0.1);
\end{scope}
\end{tikzpicture}
\ee
Then, we can take
\be
S_{a=1}^\mu=S^\mu_{i=n} \,.
\ee
The non-trivial charges under this surface are provided by the fiber $f_{in-1}^\mu$ and $f_{i=n}^\mu$, both of which have charge $\pm 2$, thus reducing the $U(1)$ generated by $S^\mu_{a=1}$ to $\Z_2$, which can be identified as the center of $\so(2n+1)$.\\
We can choose $\tilde e^\mu=e_{i=1}^\mu$.
\item For $\fg_\mu=\sp(n)$, label the nodes in the Dynkin diagram as
\be
\begin{tikzpicture}[scale=1.5]
\draw[fill=black]
(0,0) node (v1) {} circle [radius=.1]
(1,0) node (v3) {} circle [radius=.1]
(2,0) node (v4) {} circle [radius=.1]
(4,0) node (v5) {} circle [radius=.1]
(5,0) node (v6) {} circle [radius=.1];
\node (v2) at (3,0) {$\cdots$};
\draw  (v1) edge (v3);
\draw  (v3) edge (v4);
\draw  (v4) edge (v2);
\draw  (v2) edge (v5);
\node at (0,-0.3) {{$1$}};
\node at (1,-0.3) {{$2$}};
\node at (2,-0.3) {{$3$}};
\node at (4,-0.3) {{$n-1$}};
\node at (5,-0.3) {{$n$}};
\begin{scope}[shift={(-1,0)}]
\draw (5.15,0.025) -- (5.825,0.025) (5.825,-0.025) -- (5.15,-0.025);
\draw (5.6,0.1) -- (5.5,0) -- (5.6,-0.1);
\end{scope}
\end{tikzpicture}
\ee
Then, we can take
\be
S_{a=1}^\mu=\sum_{i=1}^{n}\frac{1-(-1)^i}{2}S^\mu_i \,.
\ee
Each fiber $f_{i}^\mu$ has charge $\pm 2$ under this surface, thus reducing the $U(1)$ generated by $S^\mu_{a=1}$ to $\Z_2$, which can be identified as the center of $\sp(n)$.\\
We can choose $\tilde e^\mu=e_{i=n}^\mu$.
\item For $\fg_\mu=\so(4n+2)$, label the nodes in the Dynkin diagram as
\be
\begin{tikzpicture}[scale=1.5]
\draw[fill=black]
(-0.4,0) node (v1) {} circle [radius=.1]
(0.8,0) node (v3) {} circle [radius=.1]
(2,0) node (v4) {} circle [radius=.1]
(4,0) node (v5) {} circle [radius=.1]
(5,0) node (v6) {} circle [radius=.1]
(0.8,1.2) node (v7) {} circle [radius=.1];
\node (v2) at (3,0) {$\cdots$};
\draw  (v1) edge (v3);
\draw  (v3) edge (v4);
\draw  (v4) edge (v2);
\draw  (v2) edge (v5);
\node at (-0.4,-0.3) {{$2n+1$}};
\node at (0.8,-0.3) {{$2n-1$}};
\node at (2,-0.3) {{$2n-2$}};
\node at (4,-0.3) {{$2$}};
\node at (5,-0.3) {{$1$}};
\draw  (v5) edge (v6);
\draw  (v7) edge (v3);
\node at (0.8,1.5) {{$2n$}};
\end{tikzpicture}
\ee
Then, we can take
\be
S_{a=1}^\mu=3S^\mu_{i=2n+1}+S^\mu_{i=2n}+\sum_{i=1}^{2n-1}\left(1-(-1)^i\right)S^\mu_i \,.
\ee
Each fiber $f_{i}^\mu$ has charge $\pm 4$ under this surface except for $f^\mu_{i=2n}$ which has 0 charge. Thus, the $U(1)$ generated by $S^\mu_{a=1}$ is reduced to $\Z_4$, which can be identified as the center of $\so(4n+2)$.\\
We can choose $\tilde e^\mu=e_{i=1}^\mu$.
\item For $\fg_\mu=\so(4n)$, label the nodes in the Dynkin diagram as
\be
\begin{tikzpicture}[scale=1.5]
\draw[fill=black]
(-0.4,0) node (v1) {} circle [radius=.1]
(0.8,0) node (v3) {} circle [radius=.1]
(2,0) node (v4) {} circle [radius=.1]
(4,0) node (v5) {} circle [radius=.1]
(5,0) node (v6) {} circle [radius=.1]
(0.8,1.2) node (v7) {} circle [radius=.1];
\node (v2) at (3,0) {$\cdots$};
\draw  (v1) edge (v3);
\draw  (v3) edge (v4);
\draw  (v4) edge (v2);
\draw  (v2) edge (v5);
\node at (-0.4,-0.3) {{$2n-1$}};
\node at (0.8,-0.3) {{$2n-2$}};
\node at (2,-0.3) {{$2n-3$}};
\node at (4,-0.3) {{$2$}};
\node at (5,-0.3) {{$1$}};
\draw  (v5) edge (v6);
\draw  (v7) edge (v3);
\node at (0.8,1.5) {{$2n$}};
\end{tikzpicture}
\ee
Then, we can take
\begin{align}
S_{a=1}^\mu&=\sum_{i=1}^{2n-1}\frac{1-(-1)^i}2S^\mu_i\\
S_{a=2}^\mu&=S^\mu_{i=2n}+\sum_{i=1}^{2n-2}\frac{1-(-1)^i}2S^\mu_i \,.
\end{align}
Each fiber $f_{i}^\mu$ has charge $\pm 2$ under $S_{a=1}^\mu$ except for $f^\mu_{i=2n}$ which has 0 charge. Similarly, each fiber $f_{i}^\mu$ has charge $\pm 2$ under $S_{a=2}^\mu$ except for $f^\mu_{i=2n-1}$ which has 0 charge. Thus, the $U(1)\times U(1)$ generated by $S^\mu_{a=1}$ and $S^\mu_{a=2}$ is reduced to $\Z_2\times\Z_2$, which can be identified as the center of $\so(4n)$.\\
We can choose $\tilde e^\mu=e_{i=1}^\mu$.
\item For $\fg_\mu=\fe_6$, label the nodes in the Dynkin diagram as
\be
\begin{tikzpicture}[scale=1.5]
\draw[fill=black]
(1,0) node (v1) {} circle [radius=.1]
(2,0) node (v3) {} circle [radius=.1]
(3,0) node (v4) {} circle [radius=.1]
(4,0) node (v5) {} circle [radius=.1]
(5,0) node (v6) {} circle [radius=.1]
(3,1) node (v7) {} circle [radius=.1];
\node at (1,-0.3) {{$5$}};
\node at (2,-0.3) {{$4$}};
\node at (3,-0.3) {{$3$}};
\node at (4,-0.3) {{$2$}};
\node at (5,-0.3) {{$1$}};
\draw  (v5) edge (v6);
\node at (3,1.3) {{$6$}};
\draw  (v4) edge (v5);
\draw  (v3) edge (v4);
\draw  (v1) edge (v3);
\draw  (v7) edge (v4);
\end{tikzpicture}
\ee
Then, we can take
\be
S_{a=1}^\mu=\sum_{i=1}^{5}iS^\mu_i \,.
\ee
Only the fiber $f_{i=5}^\mu$ and $f_{i=6}^\mu$ have non-trivial charges under under this surface, which are $6$ and $3$ respectively. Thus, the $U(1)$ generated by $S^\mu_{a=1}$ is reduced to $\Z_3$, which can be identified as the center of $\fe_6$.\\
We can choose $\tilde e^\mu=e_{i=1}^\mu$.
\item For $\fg_\mu=\fe_7$, label the nodes in the Dynkin diagram as
\be
\begin{tikzpicture}[scale=1.5]
\draw[fill=black]
(1,0) node (v1) {} circle [radius=.1]
(2,0) node (v3) {} circle [radius=.1]
(3,0) node (v4) {} circle [radius=.1]
(4,0) node (v5) {} circle [radius=.1]
(5,0) node (v6) {} circle [radius=.1]
(6,0) node (v8) {} circle [radius=.1]
(3,1) node (v7) {} circle [radius=.1];
\node at (1,-0.3) {{$6$}};
\node at (2,-0.3) {{$5$}};
\node at (3,-0.3) {{$4$}};
\node at (4,-0.3) {{$3$}};
\node at (5,-0.3) {{$2$}};
\node at (6,-0.3) {{$1$}};
\draw  (v5) edge (v6);
\node at (3,1.3) {{$7$}};
\draw  (v4) edge (v5);
\draw  (v3) edge (v4);
\draw  (v1) edge (v3);
\draw  (v7) edge (v4);
\draw  (v6) edge (v8);
\end{tikzpicture}
\ee
Then, we can take
\be
S_{a=1}^\mu=S_{i=1}^\mu+S_{i=3}^\mu+S_{i=7}^\mu \,.
\ee
Each fiber $f_{i}^\mu$ has charge $\pm 2$ under this surface except for $f^\mu_{i=5}$ and $f^\mu_{i=6}$, both of which have 0 charge. Thus, the $U(1)$ generated by $S^\mu_{a=1}$ is reduced to $\Z_2$, which can be identified as the center of $\fe_7$.\\
We can choose $\tilde e^\mu=e_{i=1}^\mu$.
\item For $\fg_\mu=\fe_8$, label the nodes in the Dynkin diagram as
\be
\begin{tikzpicture}[scale=1.5]
\draw[fill=black]
(1,0) node (v1) {} circle [radius=.1]
(2,0) node (v3) {} circle [radius=.1]
(3,0) node (v4) {} circle [radius=.1]
(4,0) node (v5) {} circle [radius=.1]
(5,0) node (v6) {} circle [radius=.1]
(6,0) node (v8) {} circle [radius=.1]
(7,0) node (v9) {} circle [radius=.1]
(3,1) node (v7) {} circle [radius=.1];
\node at (1,-0.3) {{$7$}};
\node at (2,-0.3) {{$6$}};
\node at (3,-0.3) {{$5$}};
\node at (4,-0.3) {{$4$}};
\node at (5,-0.3) {{$3$}};
\node at (6,-0.3) {{$2$}};
\node at (7,-0.3) {{$1$}};
\draw  (v5) edge (v6);
\node at (3,1.3) {{$8$}};
\draw  (v4) edge (v5);
\draw  (v3) edge (v4);
\draw  (v1) edge (v3);
\draw  (v7) edge (v4);
\draw  (v6) edge (v8);
\draw  (v8) edge (v9);
\end{tikzpicture}
\ee
There is no linear combination of $S_i^\mu$ under which $f_j^\mu$ have charges with gcd bigger than 1, which is consistent with the fact that the center of $\fe_8$ is trivial.\\
We can choose $\tilde e^\mu=e_{i=1}^\mu$.
\item For $\fg_\mu=\ff_4$, label the nodes in the Dynkin diagram as
\be
\begin{tikzpicture}[scale=1.5]
\draw[fill=black]
(6,0) node (v3) {} circle [radius=.1]
(3,0) node (v4) {} circle [radius=.1]
(4,0) node (v5) {} circle [radius=.1]
(5,0) node (v6) {} circle [radius=.1];
\node at (6,-0.3) {{$4$}};
\node at (3,-0.3) {{$1$}};
\node at (4,-0.3) {{$2$}};
\node at (5,-0.3) {{$3$}};
\begin{scope}[shift={(-1,0)}]
\draw (5.15,0.025) -- (5.825,0.025) (5.825,-0.025) -- (5.15,-0.025);
\draw (5.6,0.1) -- (5.5,0) -- (5.6,-0.1);
\end{scope}
\draw  (v4) edge (v5);
\draw  (v6) edge (v3);
\end{tikzpicture}
\ee
There is no linear combination of $S_i^\mu$ under which $f_j^\mu$ have charges with gcd bigger than 1, which is consistent with the fact that the center of $\ff_4$ is trivial.\\
We can choose $\tilde e^\mu=e_{i=4}^\mu$.
\item For $\fg_\mu=\fg_2$, label the nodes in the Dynkin diagram as
\be
\begin{tikzpicture}[scale=1.5]
\draw[fill=black]
(4,0) node (v5) {} circle [radius=.1]
(5,0) node (v6) {} circle [radius=.1];
\node at (4,-0.3) {{$1$}};
\node at (5,-0.3) {{$2$}};
\begin{scope}[shift={(-1,0)}]
\draw (5.15,0.025) -- (5.825,0.025) (5.825,-0.025) -- (5.15,-0.025);
\draw (5.15,0) -- (5.825,0);
\draw (5.6,0.1) -- (5.5,0) -- (5.6,-0.1);
\end{scope}
\end{tikzpicture}
\ee
There is no linear combination of $S_i^\mu$ under which $f_j^\mu$ have charges with gcd bigger than 1, which is consistent with the fact that the center of $\fg_2$ is trivial.\\
We can choose $\tilde e^\mu=e_{i=2}^\mu$.
\eit

Now that we have identified the centers $\Gamma_\mu$ of $\fg_\mu$ in terms of surfaces, it is straightforward to compute the charges of other particles under $\Gamma_\mu$. Let us first consider the effect of a (full or half) hyper charged in an irreducible representation $R$ of the gauge algebra $\fg=\oplus_\mu\fg_\mu$. The highest weight of $R$ is given by some non-negative integers $n^\mu_i$ for various $i$ and $\mu$. Then, the geometry for the gauge theory must contain a curve $C$ which satisfies
\be
-C\cdot S^\mu_i = n^\mu_i \,.
\ee
Moreover, the other curves associated to this hyper can be obtained from $C$ by subtracting $f^\nu_j$ for various $j$ and $\nu$ from it. Since $f^\nu_j$ do not screen the center $\Gamma=\prod_\mu\Gamma_\mu$ potential 1-form symmetry, the screening due to the hyper is completely captured by the charge of the curve $C$ under $\Gamma$, which can be readily computed using the data provided so far. For $\fg_\mu\neq\so(4n)$, we have a single surface responsible for generating the center which can be written as
\be
S_{a=1}^\mu=\sum_{i=1}^{r_\mu}p_i^\mu S_i^\mu \,,
\ee
from which we find that the charge of the hyper under $\Gamma_\mu$ is
\be\label{ch1}
\sum_{i=1}^{r_\mu}p_i^\mu n^\mu_i~(\text{mod}~N_\mu) \,,
\ee
where $N_\mu$ is the order of $\Gamma_\mu$. On the other hand, for $\fg_\mu=\so(4n)$, the charges under $\Gamma_\mu=\Z_2^2$ are given by
\be\label{ch2}
\left(~\sum_{i=1}^{2n-1}\frac{1-(-1)^i}2n^\mu_i~(\text{mod}~2),~n^\mu_{i=2n}+\sum_{i=1}^{2n-2}\frac{1-(-1)^i}2n^\mu_i~(\text{mod}~2)\right) \,.
\ee
Thus, we have computed the charge of an arbitrary irreducible representation $R$ under the center $\Gamma$ of a semi-simple Lie algebra $\fg$. One can use the results presented here to verify the charges tabulated in Table \ref{table}.

At this point, we have incorporated the effect of the fibers and blowups living in all the Hirzebruch surfaces $S_i$. The fibers were responsible for breaking the potential 1-form symmetry down to the center and the blowups encode the reduction of the center 1-form symmetry induced by the hypermultiplets. The only contribution left to be taken into account now come from the $e$ curves of $S_i$. These contributions themselves can be further simplified drastically since we only need to take into account a single $e$ curve for each $\mu$. This follows from the result (\ref{inst}) which states that the contribution of every $e_i^\mu$ for a fixed $\mu$ is accounted by the $\tilde e^\mu$ upto the contributions coming from fibers and blowups, but we have already accounted for the contributions from fibers and blowups. So the relevant instanton contribution can be captured by the charges of $\tilde e^\mu$ under the center $\Gamma_\nu$
\be
-\tilde e^\mu\cdot S_a^\nu \,,
\ee
where $a=1$ for $\fg_\nu\neq\so(4n)$ and $a=1,2$ for $\fg_\nu=\so(4n)$.

Let us see how these instanton contributions affect gauge theories carrying a simple gauge algebra only. Consider first pure gauge theories for which geometries were provided in \cite{Bhardwaj:2019ngx}. For a pure $\su(n)$ theory with CS level $k$ such that $0\le k<n-2$, the geometry is
\be
\begin{tikzpicture} [scale=1.9]
\node (v1) at (-4.9,-0.5) {$\mathbf{1}_{n-2-k}$};
\node (v2) at (-3.1,-0.5) {$\mathbf{2}_{n-4-k}$};
\node (v3) at (-1.3,-0.5) {$\mathbf{3}_{n-6-k}$};
\node at (-4.4,-0.4) {\scriptsize{$e$}};
\node at (-3.6,-0.4) {\scriptsize{$h$}};
\draw  (v1) edge (v2);
\draw  (v2) edge (v3);
\begin{scope}[shift={(1.8,0)}]
\node at (-4.4,-0.4) {\scriptsize{$e$}};
\node at (-3.6,-0.4) {\scriptsize{$h$}};
\end{scope}
\node (v4) at (-0.2,-0.5) {$\cdots$};
\draw  (v3) edge (v4);
\node (v5) at (1.2,-0.5) {$\mathbf{(n-1)}_{2-n-k}$};
\draw  (v4) edge (v5);
\begin{scope}[shift={(3.6,0)}]
\node at (-4.4,-0.4) {\scriptsize{$e$}};
\node at (-3.2,-0.4) {\scriptsize{$h$}};
\end{scope}
\end{tikzpicture}
\ee
where $\mathbf{i}_n$ is a notation for a Hirzebruch surface $S_i=\bF_n$ without any blowups. An edge between two surfaces denotes an intersection between the two surfaces. The labels on each end of the edge denote the gluing curves inside the two surfaces being identified to construct the intersection. We can compute
\begin{align}
-\tilde e\cdot \left(\sum_{i=1}^{n-1}iS_i\right)=-e_1\cdot S_1-2e_1\cdot S_2&=(k+4-n)+(2n-4-2k)\\
&=n-k=-k~(\text{mod}~n) \,,
\end{align}
which reproduces the contribution from the instanton proposed in \cite{Morrison:2020ool}. Similarly, for $k=n-2+2m$ with $m\ge0$ the geometry is
\be
\begin{tikzpicture} [scale=1.9]
\node (v1) at (-4.9,-0.5) {$\mathbf{1}_{0}$};
\node (v2) at (-3.1,-0.5) {$\mathbf{2}_{4-n+k}$};
\node (v3) at (-2,-0.5) {$\cdots$};
\node at (-4.5,-0.4) {\scriptsize{$e$+$mf$}};
\node at (-3.6,-0.4) {\scriptsize{$e$}};
\draw  (v1) edge (v2);
\draw  (v2) edge (v3);
\begin{scope}[shift={(1.8,0)}]
\node at (-4.4,-0.4) {\scriptsize{$h$}};
\node at (-3.2,-0.4) {\scriptsize{$e$}};
\end{scope}
\node (v4) at (-0.7,-0.5) {$\mathbf{(n-2)}_{n-4+k}$};
\draw  (v3) edge (v4);
\node (v5) at (1.3,-0.5) {$\mathbf{(n-1)}_{n-2+k}$};
\draw  (v4) edge (v5);
\begin{scope}[shift={(3.6,0)}]
\node at (-3.6,-0.4) {\scriptsize{$h$}};
\node at (-3,-0.4) {\scriptsize{$e$}};
\end{scope}
\end{tikzpicture}
\ee
from which we compute
\begin{align}
-\tilde e\cdot \left(\sum_{i=1}^{n-1}iS_i\right)=-e_1\cdot S_1-2e_1\cdot S_2&=2-2m\\
&=-k~(\text{mod}~n)
\end{align}
For $k=n-2+2m$ with $m\ge0$ the geometry is
\be
\begin{tikzpicture} [scale=1.9]
\node (v1) at (-4.9,-0.5) {$\mathbf{1}_{1}$};
\node (v2) at (-3.1,-0.5) {$\mathbf{2}_{4-n+k}$};
\node (v3) at (-2,-0.5) {$\cdots$};
\node at (-4.5,-0.4) {\scriptsize{$h$+$mf$}};
\node at (-3.6,-0.4) {\scriptsize{$e$}};
\draw  (v1) edge (v2);
\draw  (v2) edge (v3);
\begin{scope}[shift={(1.8,0)}]
\node at (-4.4,-0.4) {\scriptsize{$h$}};
\node at (-3.2,-0.4) {\scriptsize{$e$}};
\end{scope}
\node (v4) at (-0.7,-0.5) {$\mathbf{(n-2)}_{n-4+k}$};
\draw  (v3) edge (v4);
\node (v5) at (1.3,-0.5) {$\mathbf{(n-1)}_{n-2+k}$};
\draw  (v4) edge (v5);
\begin{scope}[shift={(3.6,0)}]
\node at (-3.6,-0.4) {\scriptsize{$h$}};
\node at (-3,-0.4) {\scriptsize{$e$}};
\end{scope}
\end{tikzpicture}
\ee
from which we compute
\begin{align}
-\tilde e\cdot \left(\sum_{i=1}^{n-1}iS_i\right)=-e_1\cdot S_1-2e_1\cdot S_2&=1-2m\\
&=-k~(\text{mod}~n) \,.
\end{align}
Thus we find that for pure $\su(n)$ with CS level $k$, the instanton contributions can be accounted for by considering an instanton of charge $-k~(\text{mod}~n)$ under the center $\Z_n$.

For pure $\so(2n+1)$, the geometry is
\be
\begin{tikzpicture} [scale=1.9]
\node (v1) at (-2.9,-0.5) {$\mathbf{1}_{2n-5}$};
\node (v3) at (-1.3,-0.5) {$\mathbf{2}_{2n-7}$};
\node at (-2.5,-0.4) {\scriptsize{$e$}};
\node at (-1.7,-0.4) {\scriptsize{$h$}};
\node (v4) at (-0.3,-0.5) {$\cdots$};
\draw  (v3) edge (v4);
\node (v5) at (0.9,-0.5) {$\mathbf{(n-2)}_{1}$};
\draw  (v4) edge (v5);
\begin{scope}[shift={(3.6,0)}]
\node at (-4.5,-0.4) {\scriptsize{$e$}};
\node at (-3.2,-0.4) {\scriptsize{$h$}};
\end{scope}
\draw  (v1) edge (v3);
\node (v2) at (2.4,-0.5) {$\mathbf{(n-1)}_1$};
\draw  (v5) edge (v2);
\node at (1.4,-0.4) {\scriptsize{$e$}};
\node at (1.9,-0.4) {\scriptsize{$e$}};
\node (v6) at (3.7,-0.5) {$\mathbf{n}_6$};
\draw  (v2) edge (v6);
\node at (2.9,-0.4) {\scriptsize{$2h$}};
\node at (3.4,-0.4) {\scriptsize{$e$}};
\end{tikzpicture}
\ee
from which we compute
\be
-\tilde e\cdot S_n=-e_1\cdot S_n=0 \,.
\ee
Thus the instanton associated to $\so(2n+1)$ is not charged under its center.

For pure $\sp(n)$ with $\theta=n\pi~(\text{mod}~2\pi)$, the geometry is
\be
\begin{tikzpicture} [scale=1.9]
\node (v1) at (-2.9,-0.5) {$\mathbf{1}_{2n+2}$};
\node (v3) at (-1.3,-0.5) {$\mathbf{2}_{2n}$};
\node at (-2.5,-0.4) {\scriptsize{$e$}};
\node at (-1.7,-0.4) {\scriptsize{$h$}};
\node (v4) at (-0.3,-0.5) {$\cdots$};
\draw  (v3) edge (v4);
\node (v5) at (0.9,-0.5) {$\mathbf{(n-2)}_{8}$};
\draw  (v4) edge (v5);
\begin{scope}[shift={(3.6,0)}]
\node at (-4.5,-0.4) {\scriptsize{$e$}};
\node at (-3.2,-0.4) {\scriptsize{$h$}};
\end{scope}
\draw  (v1) edge (v3);
\node (v2) at (2.4,-0.5) {$\mathbf{(n-1)}_6$};
\draw  (v5) edge (v2);
\node at (1.4,-0.4) {\scriptsize{$e$}};
\node at (1.9,-0.4) {\scriptsize{$h$}};
\node (v6) at (3.7,-0.5) {$\mathbf{n}_1$};
\draw  (v2) edge (v6);
\node at (3.4,-0.4) {\scriptsize{$2h$}};
\node at (2.9,-0.4) {\scriptsize{$e$}};
\end{tikzpicture}
\ee
from which we compute
\be
-\tilde e\cdot \left(\sum_{i=1}^{n}\frac{1-(-1)^i}{2}S_i\right)=-e_n\cdot \left(\frac{1-(-1)^n}{2}S_n\right)=\frac{1-(-1)^n}{2} \,,
\ee
which is only non-trivial for $n=2m+1$.

Similarly, for pure $\sp(n)$ with $\theta=(n+1)\pi~(\text{mod}~2\pi)$, the geometry is
\be
\begin{tikzpicture} [scale=1.9]
\node (v1) at (-2.9,-0.5) {$\mathbf{1}_{2n+2}$};
\node (v3) at (-1.3,-0.5) {$\mathbf{2}_{2n}$};
\node at (-2.5,-0.4) {\scriptsize{$e$}};
\node at (-1.7,-0.4) {\scriptsize{$h$}};
\node (v4) at (-0.3,-0.5) {$\cdots$};
\draw  (v3) edge (v4);
\node (v5) at (0.9,-0.5) {$\mathbf{(n-2)}_{8}$};
\draw  (v4) edge (v5);
\begin{scope}[shift={(3.6,0)}]
\node at (-4.5,-0.4) {\scriptsize{$e$}};
\node at (-3.2,-0.4) {\scriptsize{$h$}};
\end{scope}
\draw  (v1) edge (v3);
\node (v2) at (2.4,-0.5) {$\mathbf{(n-1)}_6$};
\draw  (v5) edge (v2);
\node at (1.4,-0.4) {\scriptsize{$e$}};
\node at (1.9,-0.4) {\scriptsize{$h$}};
\node (v6) at (3.9,-0.5) {$\mathbf{n}_0$};
\draw  (v2) edge (v6);
\node at (3.5,-0.4) {\scriptsize{$2e$+$f$}};
\node at (2.9,-0.4) {\scriptsize{$e$}};
\end{tikzpicture}
\ee
from which we compute
\be
-\tilde e\cdot \left(\sum_{i=1}^{n}\frac{1-(-1)^i}{2}S_i\right)=-\frac{1-(-1)^{n-1}}{2}~(\text{mod}~2) \,,
\ee
which is only non-trivial for $n=2m$. Thus, combining both the cases, we find that the instanton has a non-trivial contribution only for $\sp(n)$ with $\theta=\pi$ for which it contributes with charge $1$ under the center $\Z_2$ associated to $\sp(n)$. This agrees with the proposal of \cite{Morrison:2020ool}.

For pure $\so(4n+2)$ the geometry is
\be
\begin{tikzpicture} [scale=1.9]
\node (v1) at (-2.9,-0.5) {$\mathbf{1}_{4n-4}$};
\node (v3) at (-1.3,-0.5) {$\mathbf{2}_{4n-6}$};
\node at (-2.5,-0.4) {\scriptsize{$e$}};
\node at (-1.7,-0.4) {\scriptsize{$h$}};
\node (v4) at (-0.2,-0.5) {$\cdots$};
\draw  (v3) edge (v4);
\node (v5) at (0.9,-0.5) {$\mathbf{(2n-2)}_{2}$};
\draw  (v4) edge (v5);
\begin{scope}[shift={(3.6,0)}]
\node at (-4.5,-0.4) {\scriptsize{$e$}};
\node at (-3.3,-0.4) {\scriptsize{$h$}};
\end{scope}
\draw  (v1) edge (v3);
\node (v2) at (2.6,-0.5) {$\mathbf{(2n-1)}_0$};
\draw  (v5) edge (v2);
\node at (1.5,-0.4) {\scriptsize{$e$}};
\node at (2,-0.4) {\scriptsize{$e$}};
\node (v6) at (4.1,-0.5) {$\mathbf{2n}_2$};
\draw  (v2) edge (v6);
\node at (3.2,-0.4) {\scriptsize{$e$}};
\node at (3.7,-0.4) {\scriptsize{$e$}};
\node (v7) at (2.6,0.5) {$\mathbf{(2n+1)}_2$};
\draw  (v7) edge (v2);
\node at (2.5,-0.2) {\scriptsize{$e$}};
\node at (2.5,0.2) {\scriptsize{$e$}};
\end{tikzpicture}
\ee
for which we compute
\be
-\tilde e\cdot \left(3S_{i=2n+1}+S_{i=2n}+\sum_{i=1}^{2n-1}\left(1-(-1)^i\right)S_i\right)=-2e_1\cdot S_1=12-8n=0~(\text{mod}~4) \,.
\ee
Thus the instanton associated to $\so(4n+2)$ is not charged under its $\Z_4$ center.

For pure $\so(4n)$ the geometry is
\be
\begin{tikzpicture} [scale=1.9]
\node (v1) at (-2.9,-0.5) {$\mathbf{1}_{4n-6}$};
\node (v3) at (-1.3,-0.5) {$\mathbf{2}_{4n-8}$};
\node at (-2.5,-0.4) {\scriptsize{$e$}};
\node at (-1.7,-0.4) {\scriptsize{$h$}};
\node (v4) at (-0.2,-0.5) {$\cdots$};
\draw  (v3) edge (v4);
\node (v5) at (0.9,-0.5) {$\mathbf{(2n-3)}_{2}$};
\draw  (v4) edge (v5);
\begin{scope}[shift={(3.6,0)}]
\node at (-4.5,-0.4) {\scriptsize{$e$}};
\node at (-3.3,-0.4) {\scriptsize{$h$}};
\end{scope}
\draw  (v1) edge (v3);
\node (v2) at (2.6,-0.5) {$\mathbf{(2n-2)}_0$};
\draw  (v5) edge (v2);
\node at (1.5,-0.4) {\scriptsize{$e$}};
\node at (2,-0.4) {\scriptsize{$e$}};
\node (v6) at (4.1,-0.5) {$\mathbf{2n}_2$};
\draw  (v2) edge (v6);
\node at (3.2,-0.4) {\scriptsize{$e$}};
\node at (3.7,-0.4) {\scriptsize{$e$}};
\node (v7) at (2.6,0.5) {$\mathbf{(2n-1)}_2$};
\draw  (v7) edge (v2);
\node at (2.5,-0.2) {\scriptsize{$e$}};
\node at (2.5,0.2) {\scriptsize{$e$}};
\end{tikzpicture}
\ee
for which we compute
\be
-\tilde e\cdot \left(\sum_{i=1}^{2n-1}\frac{1-(-1)^i}2S_i\right)=-e_1\cdot S_1=8-4n=0~(\text{mod}~2) \,,
\ee
and
\be
-\tilde e\cdot \left(S_{i=2n}+\sum_{i=1}^{2n-2}\frac{1-(-1)^i}2S_i\right)=-e_1\cdot S_1=8-4n=0~(\text{mod}~2) \,.
\ee
Thus the instanton associated to $\so(4n)$ is not charged under its $\Z_2^2$ center.

For pure $\fe_6$ the geometry is
\be
\begin{tikzpicture} [scale=1.9]
\node at (0.6,-0.4) {\scriptsize{$h$}};
\node at (-0.1,-0.4) {\scriptsize{$e$}};
\node (v4) at (-0.4,-0.5) {$\mathbf{1}_4$};
\node (v5) at (0.9,-0.5) {$\mathbf{2}_{2}$};
\draw  (v4) edge (v5);
\node (v2) at (2.2,-0.5) {$\mathbf{3}_0$};
\draw  (v5) edge (v2);
\node at (1.2,-0.4) {\scriptsize{$e$}};
\node at (1.9,-0.4) {\scriptsize{$e$}};
\node (v6) at (3.5,-0.5) {$\mathbf{4}_2$};
\draw  (v2) edge (v6);
\node at (2.5,-0.4) {\scriptsize{$e$}};
\node at (3.2,-0.4) {\scriptsize{$e$}};
\node (v7) at (2.2,0.5) {$\mathbf{6}_2$};
\draw  (v7) edge (v2);
\node at (2.1,-0.2) {\scriptsize{$e$}};
\node at (2.1,0.2) {\scriptsize{$e$}};
\node (v8) at (4.8,-0.5) {$\mathbf{5}_4$};
\draw  (v6) edge (v8);
\node at (3.8,-0.4) {\scriptsize{$h$}};
\node at (4.5,-0.4) {\scriptsize{$e$}};
\end{tikzpicture}
\ee
for which we compute
\be
-\tilde e\cdot \left(\sum_{i=1}^{5}iS_i\right)=-e_1\cdot S_1-2e_1\cdot S_2=-2+8=0~(\text{mod}~3)
\ee
Thus the instanton associated to $\fe_6$ is not charged under its $\Z_3$ center.

For pure $\fe_7$ the geometry is
\be
\begin{tikzpicture} [scale=1.9]
\node at (0.6,-0.4) {\scriptsize{$h$}};
\node at (-0.1,-0.4) {\scriptsize{$e$}};
\node (v4) at (-0.4,-0.5) {$\mathbf{2}_4$};
\node (v5) at (0.9,-0.5) {$\mathbf{3}_{2}$};
\draw  (v4) edge (v5);
\node (v2) at (2.2,-0.5) {$\mathbf{4}_0$};
\draw  (v5) edge (v2);
\node at (1.2,-0.4) {\scriptsize{$e$}};
\node at (1.9,-0.4) {\scriptsize{$e$}};
\node (v6) at (3.5,-0.5) {$\mathbf{5}_2$};
\draw  (v2) edge (v6);
\node at (2.5,-0.4) {\scriptsize{$e$}};
\node at (3.2,-0.4) {\scriptsize{$e$}};
\node (v7) at (2.2,0.5) {$\mathbf{7}_2$};
\draw  (v7) edge (v2);
\node at (2.1,-0.2) {\scriptsize{$e$}};
\node at (2.1,0.2) {\scriptsize{$e$}};
\node (v8) at (4.8,-0.5) {$\mathbf{6}_4$};
\draw  (v6) edge (v8);
\node at (3.8,-0.4) {\scriptsize{$h$}};
\node at (4.5,-0.4) {\scriptsize{$e$}};
\node (v1) at (-1.7,-0.5) {$\mathbf{1}_6$};
\draw  (v1) edge (v4);
\node at (-0.7,-0.4) {\scriptsize{$h$}};
\node at (-1.4,-0.4) {\scriptsize{$e$}};
\end{tikzpicture}
\ee
for which we compute
\be
-\tilde e\cdot \left(S_1+S_3+S_7\right)=-e_1\cdot S_1=-4=0~(\text{mod}~2)
\ee
Thus the instanton associated to $\fe_7$ is not charged under its $\Z_2$ center.

Thus, for pure gauge theories we find that only for the case of $\su(n)$ with CS level $k$ and $\sp(n)$ with $\theta=\pi$ do we have to include contributions from instanton particles. Let us consider adding matter in the form of full hypermultiplets in some representation $R$ of $\fg$. If $\fg\neq\su(n),\sp(n)$ then the geometry for the theory can be represented as the geometry for the pure theory plus some blowups on top of the surfaces $S_i$ which are possibly glued to each other in some way \cite{Bhardwaj:2019ngx}. This means that the intersections of $\tilde e$ curve with the surfaces remain the same as in the pure case. That is, for $\fg\neq\su(n),\sp(n)$ we do not need to consider the instanton contributions.

For $\fg=\su(n)$ with CS level $k$, addition of a full hyper in a representation $R$ of $\su(n)$ shifts the CS level\footnote{Our convention for CS level differs from the convention used in \cite{Morrison:2020ool}. In our convention, CS level is defined by the tree-level contribution to the prepotential of the theory.} by $\pm\frac{A(R)}{2}$ where $A(R)$ is the \emph{anomaly coefficient} associated to $R$ (see \cite{Bhardwaj:2019ngx}). Then, for an $\su(n)$ theory with CS level $k$ and full hypers forming a (in general reducible) rep $R$, the instanton contributions can be accounted by accounting for an instanton particle of charge
\be\label{kfh}
-k+\frac{A(R)}2~(\text{mod}~n)
\ee
under the center $\Z_n$.

For $\fg=\sp(n)$, one can either add hypers such that theta angle becomes irrelevant, or add hypers such that theta angle remains relevant. If the theta angle becomes irrelevant, there are no instanton contributions to account for. If the theta angle remain relevant, then for $\theta=\pi$ we need to account for an instanton particle with charge $1~(\text{mod}~2)$ under the center $\Z_2$ of $\sp(n)$.

The above discussion wraps up the story of relevant instanton contributions for gauge theories with simple gauge algebra and matter in full hypers only. New interesting phenomena arise if we add matter in half-hypers of the simple gauge algebra. Unlike the case of full hypers discussed above, it is not possible to write a geometry carrying half-hypers in terms of geometry for the pure theory plus some blowups (that are possibly glued with each other). Thus, it is possible for the instanton contributions to be different from the instanton contributions for the pure gauge theory. As an illustrative example, consider adding a half-hyper in $\S$ to a pure $\so(12)$ gauge theory. Since the instanton contribution to the pure $\so(12)$ gauge theory is trivial, we might naively think that we only need to include the effect of matter in spinor rep $\S$, thus coming to the conclusion that the $5d$ gauge theory $\so(12)+\half\S$ has
\be
\cO=\Z_2 \,.
\ee
However, let us take a look at the geometry corresponding to this gauge theory which can be written as \cite{Bhardwaj:2020gyu}
\be
\begin{tikzpicture} [scale=1.9]
\node (v1) at (-5.3,-0.5) {$\mathbf{3}_{2}$};
\node (v2) at (-3.3,-0.5) {$\mathbf{2}_{4}$};
\node (v3) at (-9.5,-0.5) {$\mathbf{5}_{2}$};
\node (v0) at (-7.6,-0.5) {$\mathbf{4}^1_{0}$};
\draw  (v1) edge (v2);
\node at (-5.6,-0.4) {\scriptsize{$e$}};
\node at (-3,-0.8) {\scriptsize{$h$+$f$}};
\node at (-7.9,-0.4) {\scriptsize{$e$}};
\node at (-9.1,-0.4) {\scriptsize{$e$}};
\draw  (v0) edge (v1);
\node at (-7.2,-0.4) {\scriptsize{$e$}};
\node at (-5,-0.4) {\scriptsize{$h$}};
\node (v4) at (-3.3,-1.8) {$\mathbf{1}^{2}_{8}$};
\draw  (v0) edge (v3);
\draw  (v2) edge (v4);
\node at (-3.6,-0.4) {\scriptsize{$e$}};
\node at (-3.1,-1.5) {\scriptsize{$e$}};
\node at (-6.5,-0.7) {\scriptsize{$f$-$x$}};
\node at (-4.7,-0.7) {\scriptsize{$f$}};
\node at (-4.4,-1.7) {\scriptsize{$x$-$y$}};
\node at (-9.05,-0.75) {\scriptsize{$f$}};
\node at (-3.7,-1.3) {\scriptsize{$f$-$x$-$y$}};
\node at (-4.3,-1.4) {\scriptsize{$y$}};
\draw  (v3) edge (v4);
\draw  (v0) edge (v4);
\draw  (v1) edge (v4);
\node (v5) at (-7.6,0.8) {$\mathbf{6}_{1}$};
\draw  (v5) edge (v0);
\node at (-7.4,-0.2) {\scriptsize{$e$-$x$}};
\node at (-7.4,0.5) {\scriptsize{$e$}};
\end{tikzpicture}
\ee
where the notation $\mathbf{i}_n^b$ denotes a surface obtained by blowing up $b$ times a Hirzebruch surface $S_i=\bF_n$. Thus the Hirzebruch surface $S_4$ is blown up at one point and the Hirzebruch surface $S_1$ is blown up at two points where the exceptional curves associated to the two blowups are denoted as $x$ and $y$. Computing the contribution of instanton
\begin{align}
&\left(-\tilde e\cdot \left(\sum_{i=1}^{2n-1}\frac{1-(-1)^i}2S_i\right),-\tilde e\cdot \left(S_{i=2n}+\sum_{i=1}^{2n-2}\frac{1-(-1)^i}2S_i\right)\right)\\
=&\left(-e_1\cdot (S_1+S_3),-e_1\cdot(S_1+S_3)\right)\\
=&\left(-7,-7\right)\\
=&\left(1~(\text{mod}~2),1~(\text{mod}~2)\right) \,.
\end{align}
Thus we find that the instanton contribution combined with the contribution from spinor matter completely destroy the potential center 1-form symmetry of $\so(12)$ and the correct 1-form symmetry for $\so(12)+\half\S$ is
\be
\cO=\Z_1 \,.
\ee
Generalizing this, we see that for $\so(12)+n\S$ we have
\be
\cO=\Z_2 \,,
\ee
but for $\so(12)+\left(n+\half\right)\S$ we have
\be
\cO=\Z_1 \,.
\ee
A similar phenomenon occurs when we consider adding a half-hyper in $\L^3$ to an $\su(6)$ gauge theory. The geometries for this case were also discussed in \cite{Bhardwaj:2020gyu}. For CS level $k=\half-l$ with $1\le l\le 7$, the geometry can be written as
\be
\begin{tikzpicture} [scale=1.9]
\node (v1) at (-5.3,-0.5) {$\mathbf{3}_{l}$};
\node (v2) at (-3.3,-0.5) {$\mathbf{4}_{l-4}$};
\node (v3) at (-9.5,-0.5) {$\mathbf{1}_{4+l}$};
\node (v0) at (-7.6,-0.5) {$\mathbf{2}^1_{2+l}$};
\draw  (v1) edge (v2);
\node at (-5,-0.4) {\scriptsize{$e$}};
\node at (-3.8,-0.4) {\scriptsize{$h$+$f$}};
\node at (-8,-0.4) {\scriptsize{$h$}};
\node at (-9.1,-0.4) {\scriptsize{$e$}};
\draw  (v0) edge (v1);
\node at (-7.2,-0.4) {\scriptsize{$e$}};
\node at (-5.6,-0.4) {\scriptsize{$h$}};
\node (v4) at (-3.3,-1.8) {$\mathbf{5}^{2}_{l-6}$};
\draw  (v0) edge (v3);
\draw  (v2) edge (v4);
\node at (-3.1,-0.8) {\scriptsize{$e$}};
\node at (-3.1,-1.5) {\scriptsize{$h$}};
\node at (-6.5,-0.7) {\scriptsize{$f$-$x$}};
\node at (-4.7,-0.7) {\scriptsize{$f$}};
\node at (-4.4,-1.7) {\scriptsize{$x$-$y$}};
\node at (-9.1,-0.7) {\scriptsize{$f$}};
\node at (-3.7,-1.3) {\scriptsize{$f$-$x$-$y$}};
\node at (-4.3,-1.4) {\scriptsize{$y$}};
\draw  (v3) edge (v4);
\draw  (v0) edge (v4);
\draw  (v1) edge (v4);
\end{tikzpicture}
\ee
Hence the instanton contribution turns out to be
\be\label{khh}
-\tilde e\cdot \left(\sum_{i=1}^{5}iS_i\right)=-\tilde e\cdot (S_1+2S_2+5S_5)=(-2-l)+2(4+l)-5=-k+\frac32~(\text{mod}~3) \,,
\ee
where we are considering the contribution modulo 3 since the $\L^3$ matter already breaks the $\Z_6$ center down to a potential $\Z_3$ 1-form symmetry only. We obtain the same instanton contribution for other values of CS level as well, as the reader can check using the geometries presented in \cite{Bhardwaj:2020gyu}. The contribution (\ref{khh}) in the half-hyper case should be contrasted with the contribution (\ref{kfh}) in the full hyper case.

The above comments associated to matter in full vs half-hypermultiplets extend to the case of a semi-simple gauge algebra $\fg=\oplus_\mu\fg_\mu$. First of all, for the pure gauge theory based on $\fg$, the instanton $\tilde e_\mu$ has 0 charge under $\Gamma_\nu$ for $\nu\neq\mu$, and has non-trivial charge under $\Gamma_\mu$ only if $\fg_\mu=\su(n)$ or $\sp(n)_\pi$. Now, whenever there is a half-hyper charged in a mixed rep of $\fg_{\mu_1}\oplus\fg_{\mu_2}\oplus\cdots\oplus\fg_{\mu_l}\subseteq\fg$ (for $l\ge1$), there is at least one $\mu\in\{\mu_1,\mu_2,\cdots,\mu_l\}$ such that the instanton $\tilde e^\mu$ has a charge under $\Gamma_{\mu_1}\times\Gamma_{\mu_2}\times\cdots\times\Gamma_{\mu_l}$ that is different from the its charge under $\Gamma_{\mu_1}\times\Gamma_{\mu_2}\times\cdots\times\Gamma_{\mu_l}$ for the case of pure gauge theory based on $\fg$. The full hypers can again be ignored when accounting for instantonic contributions.

For example, consider an $\so(8)\oplus\su(2)$ gauge theory with a \emph{half-hyper} in bifundamental representation. Including the data of only the gauge algebras and hypermultiplet matter content, we will expect the 1-form symmetry to be
\be\label{5QO}
\cO=\Z_2\times\Z_2 \,,
\ee
but the geometry implies a $\Z_1$ 1-form symmetry as we will see below. The geometry for this theory can be written as
\be
\begin{tikzpicture} [scale=1.9]
\node (v2) at (2.2,-0.5) {$\mathbf{4}_0$};
\node (v6) at (3.7,-0.5) {$\mathbf{1}_2$};
\draw  (v2) edge (v6);
\node at (2.5,-0.4) {\scriptsize{$e$}};
\node at (3.4,-0.4) {\scriptsize{$e$}};
\node (v1) at (0.7,-0.5) {$\mathbf{3}_2$};
\node (v3) at (2.2,0.6) {$\mathbf{2}_2$};
\draw  (v1) edge (v2);
\draw  (v3) edge (v2);
\node at (1,-0.4) {\scriptsize{$e$}};
\node at (1.9,-0.4) {\scriptsize{$e$}};
\node at (2.1,-0.2) {\scriptsize{$e$}};
\node at (2.1,0.3) {\scriptsize{$e$}};
\node (v4) at (5.8,-0.5) {$\mathbf{5}^4_0$};
\draw  (v6) edge (v4);
\node at (4,-0.4) {\scriptsize{$f$}};
\node at (4.8,-0.4) {\scriptsize{$x_3$-$x_4$}};
\draw  (v3) edge (v4);
\node[rotate=-18] at (5.3427,-0.2747) {\scriptsize{$x_1$-$x_2$}};
\node at (2.5726,0.6013) {\scriptsize{$f$}};
\draw (v2) .. controls (2.2,-1) and (5.4,-1) .. (v4);
\draw (v1) .. controls (0.8,-1.2) and (5.8,-1.2) .. (v4);
\node[rotate=10] at (5.2148,-0.7235) {\scriptsize{$x_2$-$x_3$}};
\node at (2.6489,-0.7239) {\scriptsize{$f$}};
\node at (0.6209,-0.7732) {\scriptsize{$f$}};
\node at (5.95,-0.9) {\scriptsize{$f$-$x_1$-$x_2$}};
\end{tikzpicture}
\ee
From this geometry we see that the BPS instanton $e_1$ associated to $\so(8)$ has charge 1 under the center $\Z_2$ symmetry associated to $\su(2)$ (which is generated by $S_5$), and the BPS instanton $e_5$ associated to $\su(2)$ has charge $(1,0)$ under the center $\Z_2\times\Z_2$ symmetry associated to $\so(8)$ (which are generated respectively by $S_1+S_3$ and $S_1+S_2$). Out of the $\Z_2^3$ center symmetry, the blowups $x_i$ preserve a $\Z_2$ symmetry associated to $S_2+S_3$ and a $\Z_2$ symmetry associated to $S_1+S_2+S_5$. This is the $\Z_2\times\Z_2$ 1-form symmetry expected to be preserved from the field theoretic analysis. Now we need to to also consider the instantons. $e_4$ is charged as $(0,1)$ and $e_5$ is charged as $(1,0)$ under the $\Z_2^2$ symmetry preserved by the blowups. Thus, after including the contribution of instantons we find that the 1-form symmetry for $\so(8)\oplus\su(2)$ theory with a half-bifundamental is
\be\label{5QHO}
\cO=\Z_1
\ee
contrary to the expected answer (\ref{5QO}). The reader can also verify the answer (\ref{5QHO}) by directly computing the Smith normal form of the charge matrix $Q^{\alpha i}$ associated to the above geometry.

\subsection{1-form symmetry of $5d$ KK theories}
In this paper, we will use the term ``$5d$ KK theories'' to refer to $5d$ theories obtained by compactifying a $6d$ SCFT or LST on a circle of finite non-zero radius. The terminology stresses the fact that these $5d$ theories are different for standard $5d$ quantum field theories because they contain the KK mode arising from the circle compactification.

Upon compactification of a $6d$ theory on a circle, we can turn on Wilson lines in the flavor symmetry group of the $6d$ theory. For the continuous part of the flavor symmetry\footnote{Throughout this paper, we use the terms ``flavor symmetry'' and ``0-form symmetry'' interchangeably.} group, these Wilson lines become the mass parameters of the $5d$ KK theory. For the discrete part of the flavor symmetry group, the Wilson lines are discrete and hence parametrize different $5d$ KK theories. Two discrete Wilson lines related by a discrete background gauge transformation (valued in the discrete \emph{global} symmetry group) are equivalent on a circle, and hence lead to the same $5d$ KK theory. We refer to non-trivial discrete Wilson lines upto discrete background gauge transformations as \emph{twists}.

\subsubsection{Untwisted case}\label{5KKU}
Let us first consider the untwisted circle compactification of a $6d$ theory. The 1-form symmetry $\cO_{6d}$ of the $6d$ theory is generated by topological operators of codimension 2. Upon compactifying the $6d$ theory on a circle, we can either wrap these operators along the circle or insert them at a point on the circle. Wrapping these operators along the circle gives rise to 1-form symmetries in the $5d$ theory, while inserting the operators at a point gives rise to 0-form symmetries in the $5d$ theory. The $5d$ theory contains both the 1-form and 0-form symmetries descending from 1-form symmetry of the $6d$ theory.

Similarly, the 2-form symmetry $\cT_{6d}$ of the $6d$ theory is generated by topological operators of codimension 3. Wrapping the operators along a circle would give rise to 2-form symmetry in the $5d$ theory, while inserting the operators at a point gives rise to 1-form symmetries of the $5d$ theory. However, unlike the case of $\cO_{6d}$ discussed above, the $5d$ theory cannot simultaneously have both the 1-form and 2-form symmetries originating from the 2-form symmetry of the $6d$ theory. 

This is due to the fact that the 2-form symmetry of the $6d$ theory is, in a sense, ``self-dual''. That is, the $6d$ theory does not admit backgrounds for the 2-form symmetry which correspond to insertion of codimension 3 topological operators along intersecting 3-cycles. Thus, we need to choose whether we wish to keep inside the $5d$ theory the 1-form symmetry arising from the 2-form symmetry of the $6d$ theory, or the 2-form symmetry arising from the 2-form symmetry of the $6d$ theory. If we choose to keep the 1-form symmetry, then we can gauge this 1-form symmetry in the resulting $5d$ theory to obtain the $5d$ theory where we would have chosen to keep the 2-form symmetry instead, and vice-versa. In this paper, we always choose to keep the 1-form symmetry.

In conclusion, a $5d$ KK theory arising via an untwisted compactification of a $6d$ theory has 1-form symmetry group
\be\label{5KU}
\cO_{5d}=\cO_{6d}\times\cT_{6d} \,.
\ee
 
\subsubsection{Twisted case}\label{5KKT}
Discrete 0-form symmetries are generated by topological operators of codimension 1. So, we can think of a twisted KK theory as being produced by inserting, at a point of the circle, the codimension 1 topological operator associated to a discrete 0-form symmetry implementing the twist. The insertion of this topological operator results in a reduction in the 1-form symmetry of the $5d$ KK theory associated to a twisted compactification as compared to the 1-form symmetry of the $5d$ KK theory associated to the untwisted compactification of the same $6d$ theory. The reason for this reduction is that the topological operators corresponding to 0-form symmetry may act on the topological operators corresponding to the 1-form or 2-form symmetries in the $6d$ theory.

As we have discussed above, a subset of the 1-form symmetries of the $5d$ KK theory arise by wrapping the topological operators corresponding to 1-form symmetry of the $6d$ theory along the circle. In the case of a non-trivial twist, say corresponding to a discrete 0-form symmetry element $g$, we are only allowed to wrap topological operators corresponding to 1-form symmetries that are left invariant by $g$. This is because, if a topological operator corresponding to a 1-form symmetry is charged under $g$, then traversing around the circle changes the type of the topological operator as it crosses the insertion of topological operator corresponding to $g$, and hence it cannot close back to itself. The surviving 1-form symmetries form a group $\text{ker}_g(\cO_{6d})$, that is the kernel of the action of $g$ on $\cO_{6d}$.

On the other hand, another subset of the 1-form symmetries of the $5d$ KK theory arise by inserting the topological operators corresponding to 2-form symmetry of the $6d$ theory at a point on the circle. Suppose we have inserted a topological operator corresponding to a 2-form symmetry element $h$. Moving this operator around the cirle, we obtain the topological operator corresponding to the 2-form symmetry element $g\cdot h$, that is the 2-form symmetry element obtained by applying the action of $g$. Thus, as elements of the 2-form symmetry group of the $5d$ KK theory, $h$ and $g\cdot h$ are identified. More generally, since $\cT_{6d}$ is abelian, an element $h_1h_2$ of $\cT_{6d}$ is identified with the elements $g(h_1)h_2$ and $h_1g(h_2)$. This identification gives rise to an equivalence relation $\sim_g$ on $\cT_{6d}$. This means that the 1-form symmetry group of the $5d$ KK theory arising from 1-form symmetry group of the $6d$ theory is the projection $\cT_{6d}/\sim_g$.

In total, we can write the 1-form symmetry group of the $5d$ KK theory obtained by $g$-twist of a $6d$ theory as
\be\label{5KT}
\cO_{5d}=\frac{\cT_{6d}}{\sim_g}\times \text{ker}_g(\cO_{6d}) \,.
\ee

Let us discuss the structure of (\ref{5KT}) in more detail for different kinds of twists of $6d$ theories. So far these twists have been studied only in the context of $6d$ SCFTs \cite{Bhardwaj:2019fzv,Bhardwaj:2020kim} but similar structure is expected to extend to the case of $6d$ LSTs. From the study of twists of $6d$ SCFTs, we expect three different kinds of twists for $6d$ theories:
\ben
\item The first kind originate from the outer-automorphisms of the gauge algebras appearing on the tensor branch of the $6d$ theory.
\item The second kind originate from a permutation symmetry of tensor multiplets arising on the tensor branch of the $6d$ theory.
\item The third kind originate for some $6d$ theories whose tensor branch theory carries an $O(2n)$ flavor symmetry. Since $O(2n)$ has two disconnected components, the holonomies valued in the component not connected to the identity element give rise to a twisted $5d$ KK theory.
\een

Combining the twists mentioned above, one can write a general $5d$ KK theory using the following graphical notation mimicking the graphical notation used for $6d$ theories:
\be\label{GD}
\begin{tikzpicture}
\node (v1) at (-0.5,0.5) {$\Omega^{ii}$};
\node (v4) at (-0.5,1) {$\fg^{(q_i)}_i$};
\begin{scope}[shift={(2.8,0.05)}]
\node at (-0.5,0.95) {$\fg^{(q_j)}_j$};
\node (v2) at (-0.5,0.45) {$\Omega^{jj}$};
\end{scope}
\node (v3) at (0.9,0.5) {\tiny{$-\Omega^{ij}$}};
\draw (v1)--(v3);
\draw [<-](v2)--(v3);
\begin{scope}[shift={(-2.3,0.05)}]
\node (v2_1) at (-0.5,0.45) {$\Omega^{kk}$};
\end{scope}
\begin{scope}[shift={(0,1.95)}]
\node (v2_2) at (-0.5,0.45) {$\Omega^{ll}$};
\node at (-0.5,0.95) {$\fg^{(q_l)}_l$};
\end{scope}
\draw  (v2_2) edge (v4);
\draw  (v2_1) edge (v1);
\node (v5) at (-2.8,2.4) {$\left[\Z_2^{(2)}\right]$};
\draw  (v5) edge (v2_2);
\end{tikzpicture}
\ee
where each node $i$ carries a twisted or untwisted affine Lie algebra $\fg_i^{(q_i)}$. This algebra may be empty for some of the nodes, as is the case for the node $k$ in the above graph. The graph also involves the data of a \emph{non-symmetric} positive-definite integer matrix $\Omega^{ij}$ with non-positive off-diagonal entries. If $\Omega^{ij}=\Omega^{ji}$ for some specific $j\neq i$, then the nodes $j$ and $i$ are connected by $-\Omega^{ij}$ number of undirected edges, as we did in the case of $6d$ theories. We can also have directed edges which arise for example when $\Omega^{ji}=-1$ and $\Omega^{ij}<-1$. Then we join the nodes $i$ and $j$ by a directed edge pointing from $i$ to $j$ and insert a label in the middle of the edge capturing the value of $-\Omega^{ij}$. The edge between nodes $i$ and $j$ in the above graph is such an example. In addition to all of this, we can have some nodes which are attached to a $\left[\Z_2^{(2)}\right]$ which is a shorthand to denote the fact that these nodes have an $O(2n)$ flavor symmetry and we have turned on holonomies in the component disconnected to the identity. In the above graph, node $l$ is an example of such a node.

The corresponding $6d$ theory can be obtained from the graph for the $5d$ KK theory by ``unfolding'' it and removing the superscript labels $q_i$ and nodes $\left[\Z_2^{(2)}\right]$. For example, the $6d$ theory associated to the $5d$ KK theory shown in the above graph for $-\Omega^{ij}=2$ takes the following form
\be
\begin{tikzpicture}
\node (v1) at (-0.5,0.5) {$\Omega^{ii}$};
\node (v4) at (-0.5,1) {$\fg_i$};
\begin{scope}[shift={(2,0.05)}]
\node at (-0.5,0.9) {$\fg_j$};
\node (v2) at (-0.5,0.45) {$\Omega^{jj}$};
\end{scope}
\begin{scope}[shift={(-2.3,0.05)}]
\node (v2_1) at (-0.5,0.45) {$\Omega^{kk}$};
\end{scope}
\begin{scope}[shift={(0,1.95)}]
\node (v2_2) at (-0.5,0.45) {$\Omega^{ll}$};
\node at (-0.5,0.9) {$\fg_l$};
\end{scope}
\draw  (v2_2) edge (v4);
\draw  (v2_1) edge (v1);
\begin{scope}[shift={(0,-1.75)}]
\node (v3) at (-0.5,0.9) {$\fg_m$};
\node (v2_3) at (-0.5,0.45) {$\Omega^{mm}$};
\end{scope}
\draw  (v1) edge (v2);
\draw  (v1) edge (v3);
\end{tikzpicture}
\ee
with $\fg_m=\fg_j$ and $\Omega^{mm}=\Omega^{jj}$. The twist converting the above $6d$ theory to the above $5d$ KK theory contains  outer-automorphisms of $\fg_i$ and $\fg_l$ of order $q_i$ and $q_l$ respectively. This includes the possibility of no outer-automorphism twist for $\fg_i$ (or $\fg_l$) which is associated to $q_i=1$ and corresponds to the untwisted affine Lie algebra $\fg_i^{(1)}$ which is defined for any $\fg_i$. The twist also contains a permutation exchanging the tensor multiplets $m$ and $j$ which identifies $\fg_j$ and $\fg_m$ and it is also possible to have an outer-automorphism of order $q_j$ of the algebra $\fg_j$ after accounting for the identification. This identification of $j$ and $m$ induces a ``folding'' of the graph which is represented by a directed edge from $i$ to $j$ in the graph for the $5d$ KK theory. The label $-\Omega^{ij}=2$ in the middle of the directed edge tells us that the folding has been obtained by identifying 2 different nodes. Similarly, if we were to identify 3 nodes of a $6d$ SCFT, the $5d$ KK theory will contain a directed edge with a label 3 placed in the middle of the directed edge. As discussed above, the twist also includes turning on holonomies in the component disconnected to identity of the flavor symmetry $O(2n)$ associated to node $l$.

\subsubsection{Geometric analysis}\label{5KG}
We now turn to the determination of 1-form symmetry group of a $5d$ KK theory by using its M-theory geometric construction. Such geometric constructions have been extensively studied in \cite{Bhardwaj:2019fzv,Bhardwaj:2020kim,Bhardwaj:2018vuu,Bhardwaj:2018yhy,DelZotto:2017pti,Jefferson:2018irk, Apruzzi:2019opn, Apruzzi:2019vpe, Apruzzi:2019syw, Eckhard:2020jyr}. The M-theory geometric construction for a $5d$ KK theory can be easily described in terms of its graphical data of the form (\ref{GD}). For every node $i$, we have a collection of irreducible Hirzebruch surfaces (carrying some blowups) $S_{a,i}$ in the geometry. Let us first consider the nodes $i$ for which $\fg_i$ is non-trivial. The number of surfaces for each $i$ equal $r_i+1$ where $r_i$ is the rank of the gauge algebra $\fh_i$ left invariant by the outer-automorphism $\cO^{(q_i)}$ acting\footnote{For any $\cO^{(1)}$ we can choose the trivial automorphism which does not act on the gauge algebra and hence $\fh_i=\fg_i$, which makes sense since $\cO^{(1)}$ means that we do not involve any outer-automorphism twist. In this paper, we choose outer-automorphisms $\cO^{(q_i)}$ for $q_i>1$ such that the invariant gauge algebras are as follows. $\cO^{(2)}$ acting on $\su(n)$ leaves $\sp(n)$ invariant, $\cO^{(2)}$ acting on $\so(2n)$ leaves $\so(2n-1)$ invariant, $\cO^{(2)}$ acting on $\fe_6$ leaves $\ff_4$ invariant, and $\cO^{(3)}$ acting on $\so(8)$ leaves $\fg_2$ invariant.} on $\fg_i$. Let $f_{a,i}$ denote the fibers of these Hirzebruch surfaces. Then, the intersection numbers
\be
M_{ab,i}:=-f_{a,i}\cdot S_{b,i}
\ee
form the Cartan matrix of the affine Lie algebra $\fg_i^{(q_i)}$ (see \cite{Bhardwaj:2019fzv} for more details). We let $S_{0,i}$ be the surface corresponding to the affine node of the Dynkin diagram of $\fg_i^{(q_i)}$ such that $M_{ab,i}$ for $a,b\neq0$ form the Cartan matrix of $\fh_i$.

Now let us consider the nodes $i$ for which $\fg_i$ is trivial. For these nodes, there is only a single corresponding surface $S_{0,i}$ which can only be one of the following three types: $\bF_1^8$; $\bF_0^2$ with $e-x_1$ glued to $e-x_2$; or $\bF_1^2$ with the two blowups glued. For $\bF_1^8$ we define $f_{0,i}$ to be $2e+3f-\sum x_i$. For $\bF_0^2$ with $e-x_1$ glued to $e-x_2$, we define $f_{0,i}$ to be $f$. For $\bF_1^2$ with glued blowups, we define $f_{0,i}$ to be $2e+3f-2\sum x_i$.

For any two nodes $i\neq j$, we have
\be\label{off}
-f_{a,i}\cdot S_{b,j}=0 \,.
\ee
To the nodes of the Dynkin diagram of an affine Lie algebra, we can associate Coxeter labels, which are minimal positive integers that form a row null vector for the Cartan matrix of the affine Lie algebra. Similarly, we can associate dual Coxeter labels, which are minimal positive integers that form a column null vector for the Cartan matrix of the affine Lie algebra. Let us denote the Coxeter and dual Coxeter labels for $\fg_i^{(q_i)}$ by $d_{a,i}$ and $d^\vee_{a,i}$ respectively. For $\fg_i$ trivial, we let $d_{0,i}=d^\vee_{0,i}=1$. Then, to each $i$, we can assign a linear combination $S_i$ of surfaces $S_{a,i}$
\be
S_i:=\sum_a d^\vee_{a,i}S_{a,i} \,,
\ee
which has the special properties that
\be\label{f}
f_{a,i}\cdot S_i =0 \,,
\ee
and
\be\label{x}
x\cdot S_i=0 \,,
\ee
for any blowup $x$ living in any of the surfaces $S_{b,j}$. Note that we can use (\ref{off}) to write (\ref{f}) in the following more generalized form
\be\label{f'}
f_{b,j}\cdot S_i=0 \,,
\ee
for arbitrary $a,i,j$. The equations (\ref{x}) and (\ref{f'}) imply that the surfaces $S_i$ are ``null'' in the sense that all the fibers and blowups of all the Hirzebruch surfaces have no intersection with $S_i$. Note that the $e$ curves of the Hirzebruch surfaces can still intersect the null surfaces $S_i$, so it is not \emph{strictly null}.

In the last subsection, for a collection of Hirzebruch surfaces with intersection matrix describing a simple Lie algebra $\fg$, we associated the $e$ curve of a particular Hirzebruch surface to $\fg$. This curve was denoted as $\tilde e$ and it is supposed to capture the contributions of BPS instantons of $\fg$ to the breaking of 1-form symmetry. We use this fact to assign a curve $\tilde e_i$ to each $i$ as follows. For nodes $i$ with $\fg_i$ non-trivial, the surfaces $S_{a,i}$ for $a\neq 0$ and fixed $i$ form a collection of surfaces with intersection matrix describing the simple Lie algebra $\fh_i$, and we denote the $\tilde e$ curve associated to $\fh_i$ as $\tilde e_i$. For nodes $i$ with $\fg_i$ trivial, we let $\tilde e_i$ be the $e$ curves of the three possibilities discussed above. Then, it turns out that
\be\label{5KB}
-S_i\cdot \tilde e_j=\Omega^{ij} \,,
\ee
where $\Omega^{ij}$ is the matrix associated to the $5d$ KK theory as discussed above.

Now we can describe how the 1-form symmetry (\ref{5KT}) of the $5d$ KK theory is encoded in this geometry. First, we can change the basis of surfaces for each $i$ from $S_{a,i}$ to $S_i,S_{a\neq0,i}$ which is an acceptable change of basis since $d^\vee_{0,i}=1$ for any $\fg^{(q_i)}_i$. Then, we claim that the $\frac{\cT_{6d}}{\sim_g}$ part of (\ref{5KT}) is encoded in the surfaces $S_i$. Indeed $S_i$ give rise to the $\u(1)$ gauge algebras descending from KK reduction of $6d$ tensor multiplets. One can view the curves $\tilde e_i$ as BPS particles arising by wrapping (on the compactification circle) the BPS string corresponding to node $i$ in the $6d$ theory. From the above recounted facts about intersections of $S_i$ with various curves, we see that it is only the $\tilde e_i$ i.e. the BPS strings that screen the $U(1)^s$ potential 1-form symmetry generated by the surfaces $S_i$, which makes sense since $\frac{\cT_{6d}}{\sim_g}$ part of (\ref{5KT}) captures the data of the 2-form symmetry of the $6d$ theory. According to (\ref{5KB}), we find that
\be\label{5K2}
\frac{\cT_{6d}}{\sim_g}=\text{Tors}\left(\frac{\Z^s}{[\Omega^{ij}]\cdot \Z^{s}}\right) \,,
\ee
where Tors denotes the torsional part of the quotient lattice. The appearance of Tors is relevant only if the $5d$ KK theory arises via a compactification of a $6d$ LST in which case a $\u(1)$ generated by a linear combination of the $S_i$ is non-dynamical, whose contribution should be modded out, just as in the case of computation of 2-form symmetry of $6d$ LSTs discussed earlier in this paper. Just like in the case of 2-form symmetry of LSTs, the contribution from this non-dynamical $\u(1)$ gives rise to a free part in the quotient lattice, and hence we retain only the torsional part of the quotient lattice. If we specialize (\ref{5K2}) to the case of a $5d$ KK theory arising via an \emph{untwisted} compactification of a $6d$ theory, we obtain
\be
\cT_{6d}=\text{Tors}\left(\frac{\Z^s}{[\Omega^{ij}]\cdot \Z^{s}}\right) \,,
\ee
where $s$ is now captures the number of nodes in the graph associated to the $6d$ theory itself and $\Omega^{ij}$ is the matrix associated to the $6d$ theory that we discussed in Section \ref{6}. The above equation simply recovers the result of Section \ref{6T}.

The part $\text{ker}_g(\cO_{6d})$ of (\ref{5KT}) is encoded in the surfaces $S_{a\neq0,i}$. The fibers and blowups living in these surfaces give rise to a $5d$ non-abelian gauge theory $\fT$ with gauge algebra $\oplus_i\fh_i$, where the sum over $i$ is only taken over nodes with non-trivial $\fg_i$. Additional matter content for this $5d$ non-abelian gauge theory $\fT$ arises from blowups living in the surfaces $S_{0,i}$ for the nodes $i$ with $\fg_i$ non-trivial. As we have discussed in great detail in Section \ref{5GG}, the analysis of 1-form symmetries associated to the surfaces giving rise to $\oplus_i\fh_i$ can be reduced to some linear combinations of surfaces for each $i$ which capture the center symmetry $\Gamma_i$ of $\fh_i$. Potentially these surfaces give rise to a $\Gamma:=\prod_i\Gamma_i$ 1-form symmetry, which is broken according to the matter content for $\fT$ descending from the $5d$ KK theory. As discussed in Section \ref{5GG}, further breaking of $\Gamma$ is induced by instantons $\tilde e_i$ for each $\fh_i$. These curves capture precisely the BPS instanton strings associated to $\fg_i$ in the $6d$ theory as we discussed above. Following the discussion of Section \ref{5GG}, one can easily determine the charges of $\tilde e_i$ under $\Gamma$. Moreover one also needs to account for the charges of $\tilde e_i$ associated to nodes with $\fg_i$ trivial under $\Gamma$, which can be easily determined from the data of the geometry of the $5d$ KK theory. These contributions to the breaking of potential 1-form symmetry are interpreted as contributions from non-gauge-theoretic BPS strings of the $6d$ theory.

The above contributions are an end of the story if the $5d$ KK theory under consideration arises as an untwisted compactification. However, in the case of twisted compactification, one needs to consider another contribution in some cases. This contribution arises from the charge of $f_{0,i}$ under $=\Gamma_i$. The reason this is unimportant for untwisted cases is because of the fact that the genus-one fiber
\be
f_i:=\sum_a d_{a,i} f_{a,i}
\ee
has the property that
\be
f_i\cdot S_{a,i}=0
\ee
for all $a$. Since $f_{a,i}$ for $a\neq 0$ have zero charge under $\Gamma_i$, $f_{0,i}$ must have zero charge under $\Gamma_i$ as long as $d_{0,i}=1$. The latter condition is only true if the affine gauge algebra $\fg_i^{(q_i)}$ for node $i$ is untwisted, i.e. $q_i=1$. When non-trivial twist is involved, it can happen that $d_{0,i}>1$, in which case we need to include the charge of $f_{0,i}$ under $\Gamma_i$ separately into consideration. Note that we do not need to consider the charge of $f_{0,i}$ under $\Gamma_j$ for $j\neq i$ due to the fact (\ref{off}).

Thus, in conclusion, $\text{ker}_g(\cO_{6d})$ part of (\ref{5KT}) is comprised of those elements of $\Gamma$ that leave the matter content charged under $\fh$ and the extra BPS particles $\tilde e_i,f_{0,i}$ invariant.

Specializing the above discussion to the case of a $5d$ KK theory arising from an untwisted compactification of a $6d$ theory provides us with a method for computing the 1-form symmetry group $\cO_{6d}$ of the $6d$ theory itself. In this case, the $5d$ gauge theory $\fT$ is identified with the $6d$ gauge theory arising on the tensor branch of the $6d$ theory. The curves $\tilde e_i$ are in one-to-one correspondence with the BPS strings of the $6d$ theory. If $\fg_i$ is non-trivial, then the associated $\tilde e_i$ corresponds to the BPS instanton string for $\fg_i$. If $\fg_i$ is trivial, then the associated $\tilde e_i$ corresponds to the non-gauge-theoretic BPS string associated to the node $i$. The charges of $\tilde e_i$ under the center $\Gamma$ of $\fh$ are identified with the charges of the BPS strings of the $6d$ theory under the center $\Gamma$ of the $6d$ gauge algebra $\fg$. Moreover, according to the discussion of Section \ref{5GG}, we need to consider contributions of the charges of $\tilde e_i$ for non-trivial $\fg_i$ only if there are half-hypers involved or if $\fg_i=\su(n),\sp(n)$. In fact, we do not need even need to consider the case of $\fg_i=\su(n)$ since the contribution of the instanton string in this case is always accounted for by the hypermultiplet spectrum. This can be easily checked for all the possible $\fg_i=\su(n)$ that can arise in the context of $6d$ SCFTs and LSTs by taking into account (\ref{kfh}) and (\ref{khh}) along with the fact that the CS level for a $5d$ $\su(n)$ descending from a $6d$ $\su(n)$ via an untwisted compactification is always 0.\\
For example, consider the case of $\fg_i=\su(n)$ and $\Omega^{ii}=1$ such that the matter content charged under $\su(n)$ is $\L^2+(n+8)\F$. Then (\ref{kfh}) implies that the instanton string has charge $2~(\text{mod}~n)$ under the center $\Z_n$ of $\su(n)$. But since we already have a hyper in $\L^2$, as long as this hyper is not gauged by some other gauge algebra $\fg_j$, this hyper breaks the $\Z_n$ center down to $\Z_2$ and thus the charge of instanton string is irrelevant. On the other hand, remaining in the realm of $6d$ SCFTs and LSTs, it is not possible to gauge the $\L^2$ in such a way that we would be forced to account for the charge of the instanton string.\\
Thus, the only situations where the contribution of a BPS string associated to node $i$ of a $6d$ theory is relevant are as follows:
\ben
\item There is a half-hyper transforming in a mixed representation $\fg_{\mu_1}\oplus\fg_{\mu_2}\oplus\cdots\oplus\fg_{\mu_l}\subseteq\fg$ (for $l\ge1$) where $\mu_1=i$.
\item $\fg_i$ is trivial.
\item $\fg_i=\sp(n)$ with $\theta=\pi$.
\een
This justifies the claims of Section \ref{6O}.

\subsubsection{Examples}
In this subsection, we discuss examples of $5d$ KK theories arising via non-trivial twisted compactifications of $6d$ SCFTs, and discuss their 1-form symmetry using the geometric methods discussed above. We do not pursue $5d$ KK theories arising via untwisted compactifications as the computation in that case reduces to the computations performed in Section \ref{1ex}.

\ni\ubf{Example 1}: Consider the $5d$ KK theory
\be
\begin{tikzpicture}
\node (v2) at (-0.5,0.45) {$3$};
\node at (-0.45,0.95) {$\su(3)^{(2)}$};
\end{tikzpicture} \,,
\ee
which is obtained by performing an outer-automorphism twist on the $6d$ SCFT
\be
\begin{tikzpicture}
\node (v2) at (-0.5,0.45) {$3$};
\node at (-0.45,0.9) {$\su(3)$};
\end{tikzpicture} \,.
\ee
The 2-form $\Z_3$ symmetry of the $6d$ SCFT is left unaffected by the twist, and hence we expect to obtain a $\Z_3$ factor in the 1-form symmetry of the $5d$ KK theory. On the other hand, the outer automorphism twist acts on the 1-form $\Z_3$ symmetry of the $6d$ SCFT by complex conjugation, and hence we expect no contribution to the 1-form symmetry of the $5d$ KK theory from the 1-form symmetry of the $6d$ SCFT. In total, we expect that the $5d$ KK theory has
\be\label{er1}
\cO_{5d}=\Z_3 \,.
\ee

Let us verify these expectations geometrically. The geometry for the $5d$ KK theory is
\be
\begin{tikzpicture} [scale=1.9]
\node (v1) at (-3.05,-0.5) {$\mathbf{0}_{10}$};
\node (v2) at (-1.4,-0.5) {$\mathbf{1}_0$};
\draw  (v1) edge (v2);
\node at (-2.75,-0.4) {\scriptsize{$e$}};
\node at (-1.8,-0.4) {\scriptsize{$4e$+$f$}};
\end{tikzpicture}
\ee
We claimed above that the contribution to the 1-form symmetry of $5d$ KK theory from the 2-form symmetry of the $6d$ SCFT can be computed by finding the Smith normal form for $\Omega^{ij}$ associated to the $5d$ KK theory where $\Omega^{ij}$ can be computed geometrically via (\ref{5KB}). For the above geometry there is a single index $i$, and we have
\be
S_i=S_0+2S_1
\ee
and
\be
\tilde e_i=e_1 \,.
\ee
We can compute
\be
\Omega^{ii}=-S_i\cdot \tilde e_i=-(S_0+2S_1)\cdot e_1=-(4e_1+f_1)\cdot e_1-2K_1\cdot e_1=-1+4=3
\ee
which indeed is precisely what we expect. And hence we find that the 2-form part of the $6d$ theory indeed contributes $\Z_3$ factor to the 1-form symmetry of the $5d$ theory.

To compute the contribution to the 1-form symmetry of $5d$ KK theory from the 1-form symmetry of the $6d$ SCFT, we need to first delete the surface $S_0$ leaving us with the geometry
\be
\begin{tikzpicture} [scale=1.9]
\node (v2) at (-1.4,-0.5) {$\mathbf{1}_0$};
\end{tikzpicture}
\ee
which gives rise to a $5d$ non-abelian gauge theory $\fT=\su(2)$ without any matter. Note that there is no extra matter content coming from $S_0$ since $S_0$ contains no blowups. The potential center 1-form symmetry associated to $\fT$ is $\Z_2$ spanned by the surface $S_1$. Under this, we see that $\tilde e_i$ has charge
\be
-e_1\cdot S_1=-e_1\cdot K_1=2=0~(\text{mod}~2)
\ee
and $f_0$ has charge
\be
-f_0\cdot S_1=-f_0\cdot e_0=-1=1~(\text{mod}~2)
\ee
implying that the $\Z_2$ center is broken, and thus there is no contribution to the 1-form symmetry of $5d$ KK theory from the 1-form symmetry of the $6d$ SCFT, confirming the expected result (\ref{er1}).

\vspace{8pt}

\ni\ubf{Example 2}: Consider the $5d$ KK theory
\be
\begin{tikzpicture}
\node (v1) at (-0.5,0.4) {2};
\begin{scope}[shift={(1.7,-0.05)}]
\node (v2) at (-0.5,0.45) {$2$};
\end{scope}
\node (v6) at (0.4,0.4) {\tiny{2}};
\draw [<-] (v1) edge (v6);
\draw  (v6) edge (v2);
\end{tikzpicture}
\ee
which carries non non-trivial gauge algebra. This KK theory is obtained by applying a permutation twist on the following $6d$ SCFT
\be
\begin{tikzpicture}
\node (v1) at (-0.5,0.4) {2};
\begin{scope}[shift={(1.7,-0.05)}]
\node (v2) at (-0.5,0.45) {$2$};
\end{scope}
\begin{scope}[shift={(3.4,-0.05)}]
\node (v3) at (-0.5,0.45) {$2$};
\end{scope}
\draw  (v1) edge (v2);
\draw  (v2) edge (v3);
\end{tikzpicture}
\ee
which is the $A_3$ $\cN=(2,0)$ theory. As such it has a $\Z_4$ 2-form symmetry which is acted upon by the permutation twist. The $\Z_4$ can be identified as the center of $A_3$ and the permutation can be identified as the outer-automorphism of $A_3$ Lie algebra which acts by a complex conjugation on the center $\Z_4$ when $\Z_4$ is viewed as a subgroup of $U(1)$. The complex conjugation leaves only the $\Z_2$ subgroup of $\Z_4$ invariant, and hence we expect the $5d$ KK theory to attain a $\Z_2$ 1-form symmetry factor descending from the $\Z_4$ 2-form symmetry of the $6d$ SCFT. On the other hand, the $6d$ theory has no 1-form symmetry, and hence we expect the full 1-form symmetry of the $5d$ KK theory to be
\be\label{er2}
\cO_{5d}=\Z_2 \,.
\ee

Let us verify this geometrically. The geometry for the $5d$ KK theory can be written as
\be
\begin{tikzpicture} [scale=1.9]
\node (v8) at (2.7,-2) {$\mathbf{1}^{1+1}_{0}$};
\node (v7_1) at (4.4,-2) {$\mathbf{2}^{1+1}_{0}$};
\node at (3.95,-1.9) {\scriptsize{$f$-$y,y$}};
\node at (3.2,-1.9) {\scriptsize{$2f$-$x,x$}};
\node (v3_1) at (3.6,-2) {\scriptsize{2}};
\draw  (v8) edge (v3_1);
\draw  (v3_1) edge (v7_1);
\draw (v8) .. controls (2.1,-2.7) and (3.3,-2.7) .. (v8);
\node at (2.3,-2.3) {\scriptsize{$e$-$x$}};
\node at (3.1,-2.3) {\scriptsize{$e$-$y$}};
\draw (v7_1) .. controls (3.8,-2.7) and (5,-2.7) .. (v7_1);
\node at (4,-2.3) {\scriptsize{$e$-$x$}};
\node at (4.8,-2.3) {\scriptsize{$e$-$y$}};
\end{tikzpicture}
\ee
where we label the two nodes by $i$ and $j$. We have $S_i=S_{0,i}=S_1$ and $S_j=S_{0,j}=S_2$. Moreover, $\tilde e_i= e_1$ and $\tilde e_j=e_2$. We can compute the matrix $-S_i\cdot\tilde e_j$ to be
\be
\begin{pmatrix}
2&-1\\
-2&2
\end{pmatrix} \,.
\ee
which is indeed the matrix associated to the graph of the $5d$ KK theory. Computing its Smith normal form indeed reveals a $\Z_2$ contribution to the 1-form symmetry of the $5d$ KK theory. On the other hand, since both surfaces are affine surfaces, deleting them, leads to a trivial theory with no center, and hence there is no other contribution to the 1-form symmetry of the $5d$ KK theory, and we have recovered the expected result (\ref{er2}).

\subsection{Brane-web and GTP Analysis}

A subclass of 5d SCFTs have a description in terms of 5-brane webs \cite{Aharony:1997bh}, or dually in terms of generalized toric diagrams (GTP, or dot diagrams) \cite{Benini:2009gi}. We now discuss how the 1-form symmetry is encoded in this formulation of the theories, in particular how the IR gauge theory description, by inclusion of the instanton particles gives rise to the correct UV higher form symmetry. 
For models that are toric, it was argued in \cite{Morrison:2020ool}, that the 1-form symmetry of the 5d SCFT realized in terms of a toric 
fan $\{\mathbf{v}_i\}$, $i=1, \cdots, f+3$, $f$= rank of the flavor group, and with $\mathbf{v}_i = (v_i^1 , v_i^2, 1)\in \mathbb{Z}^3$, then 
\be\label{ToricO}
\mathcal{O} = \mathbb{Z}_{a_1} \oplus   \mathbb{Z}_{a_2}\oplus  \mathbb{Z}_{a_3}\,,
\ee  
with 
\be
\text{diag} (a_1, a_2, a_3) =  \text{SNF} (\mathbf{v}_1 \cdots \mathbf{v}_{f+3}) \,,
\ee
where $\text{SNF}$ is the Smith normal form, applied to the matrix of vectors in the fan. 
This is entirely independent on the resolution data and therefore computes the 1-form symmetry of the SCFT. 

In the dual web, this corresponds to taking the SNF for the $(p,q)$-charges of the external 5-branes 
\be\label{WebOne}
\text{diag} (n_1, n_2, n_3) =  \text{SNF}\left( \begin{matrix} p_1 & q_1 \cr\vdots &\vdots \cr  p_{f+3} & q_{f+3} \end{matrix}\right) \,,
\ee
When an IR gauge theory description exists, the naive expectation from the gauge theory can be that the 1-form symmetry is larger than the one of the SCFT. However as we have argued the instanton particles can be charged under the 1-form symmetry and thereby correct the classical expectation. The resulting 1-form symmetry is then always in agreement with that of the SCFT.  
We exemplify this in the case pure $SU(N)_k$. Field-theoretically we know that the 1-form symmetry is
\be
\cO= \Z_{\text{gcd}(N, k)} \,.
\ee
For pure $SU(N)_0$ the toric diagram is (shown here for $N=4$)
\be
\begin{tikzpicture}[x=.5cm,y=.5cm]
\draw[step=.5cm,gray,very thin] (0,0) grid (2,4);
\draw[ligne] (0,0)--(1,0)--(2,4)--(1,4)-- (0,0); 
\draw[ligne] (1,0)--(1,1) -- (1,2)--(1,3)-- (1,4) ;
\node[bd] at (0,0) {}; 
\node[bd] at (1,0) {}; 
\node[bd] at (2,4) {}; 
\node[bd] at (1,4) {}; 
\node[bd] at (1,1) {}; 
\node[bd] at (1,2) {}; 
\node[bd] at (1,3) {}; 
\node[bd] at (1,4) {}; 
\end{tikzpicture} \,,
\ee
One can compute using the above prescription that the 1-form symmetry associated to the above toric diagram is $\Z_4$. On the other hand, consider  pure $SU(N)_1$ for which the toric diagram is (shown here for $N=4$): 
\be
\begin{tikzpicture}[x=.5cm,y=.5cm]
\draw[step=.5cm,gray,very thin] (0,0) grid (2,4);
\draw[ligne] (0,0)--(1,0)--(2,3)--(1,4)-- (0,0); 
\draw[ligne] (1,0)--(1,1) -- (1,2)--(1,3)-- (1,4) ;
\node[bd] at (0,0) {}; 
\node[bd] at (1,0) {}; 
\node[bd] at (2,3) {}; 
\node[bd] at (1,4) {}; 
\node[bd] at (1,1) {}; 
\node[bd] at (1,2) {}; 
\node[bd] at (1,3) {}; 
\node[bd] at (1,4) {}; 
\end{tikzpicture} \,,
\ee
Computing the 1-form symmetry using the above prescription we find that $\cO=\Z_1$. If we delete either the right-most or the left-most black dot then computing SNF leads to $\Z_4$. This implies that the left-most and right-most black dots capture the instanton contribution. Indeed, this fact was already observed in \cite{Closset:2018bjz}, see also related observations in \cite{Albertini:2020mdx}.

Here we conjecture that there is a generalization to non-toric, generalized toric polygon (GTP). Consider a GTP, comprised of black and white vertices, and bring it into a convex form (see \cite{vanBeest:2020kou}). The 1-form symmetry is computed in the same way as (\ref{ToricO}), except we include all vertices that lie on the polygon -- i.e. all white dots get converted into black dots. 
The conjecture is that the resulting toric polygon has the same 1-form symmetry as the diagram with white dots.

Consider e.g. $\mathfrak{su}(4)_0 +\L^2$, whose GTP is the left diagram
\be
\begin{tikzpicture}[x=.5cm,y=.5cm]
\draw[step=.5cm,gray,very thin] (0,-2) grid (3,2);
\draw[ligne] (0,0)--(1,-1)--(2,-2)--(3,-1)-- (2,2)-- (1,1) --(0,0); 
\node[bd] at (0,0) {}; 
\node[wd] at (1,-1) {}; 
\node[bd] at (2,-2) {}; 
\node[bd] at (3,-1) {}; 
\node[bd] at (2,2) {}; 
\node[bd] at (1,1) {}; 
\end{tikzpicture} \qquad \qquad 
\begin{tikzpicture}[x=.5cm,y=.5cm]
\draw[step=.5cm,gray,very thin] (0,-2) grid (3,2);
\draw[ligne] (0,0)--(1,-1)--(2,-2)--(3,-1)-- (2,2)-- (1,1) --(0,0); 
\node[bd] at (0,0) {}; 
\node[bd] at (1,-1) {}; 
\node[bd] at (2,-2) {}; 
\node[bd] at (3,-1) {}; 
\node[bd] at (2,2) {}; 
\node[bd] at (1,1) {}; 
\end{tikzpicture} \,,
\ee
Computing the 1-form symmetry from the right diagram results in 
\be
\mathcal{O} = \mathbb{Z}_2 \,.
\ee
The right hand GTP describes an $\su(2)_0\oplus\su(4)_0$ gauge theory carrying a bifundamental
which indeed has the same 1-form symmetry.

Similarly for $\mathfrak{su}(6)_0 + \mathbf{AS}$, which has GTP given by the left diagram of 
\be
\begin{tikzpicture}[x=.5cm,y=.5cm]
\draw[step=.5cm,gray,very thin] (0,-3) grid (4,3);
\draw[ligne] (0,0)--(1,-1)--(2,-2)--(3,-3)--(4,-1)--(3,3)-- (2,2)--(1,1)-- (0,0); 
\node[bd] at (0,0) {}; 
\node[wd] at (1,-1) {}; 
\node[wd] at (2,-2) {}; 
\node[wd] at (3,-3) {}; 
\node[bd] at (4,-1) {}; 
\node[bd] at (3,3) {}; 
\node[wd] at (2,2) {}; 
\node[bd] at (1,1) {}; 
\end{tikzpicture} \qquad \qquad 
\begin{tikzpicture}[x=.5cm,y=.5cm]
\draw[step=.5cm,gray,very thin] (0,-3) grid (4,3);
\draw[ligne] (0,0)--(1,-1)--(2,-2)--(3,-3)--(4,-1)--(3,3)-- (2,2)--(1,1)-- (0,0); 
\node[bd] at (0,0) {}; 
\node[bd] at (1,-1) {}; 
\node[bd] at (2,-2) {}; 
\node[bd] at (3,-3) {}; 
\node[bd] at (4,-1) {}; 
\node[bd] at (3,3) {}; 
\node[bd] at (2,2) {}; 
\node[bd] at (1,1) {}; 
\end{tikzpicture} \,.
\ee
The right hand diagram is $\mathfrak{su}(2)_0-\mathfrak{su}(4)_0-\mathfrak{su}(6)_0$. Indeed both theories have 
$\mathcal{O}=\mathbb{Z}_2$ 1-form symmetry.

\begin{figure}
\centering
\includegraphics[width=12cm]{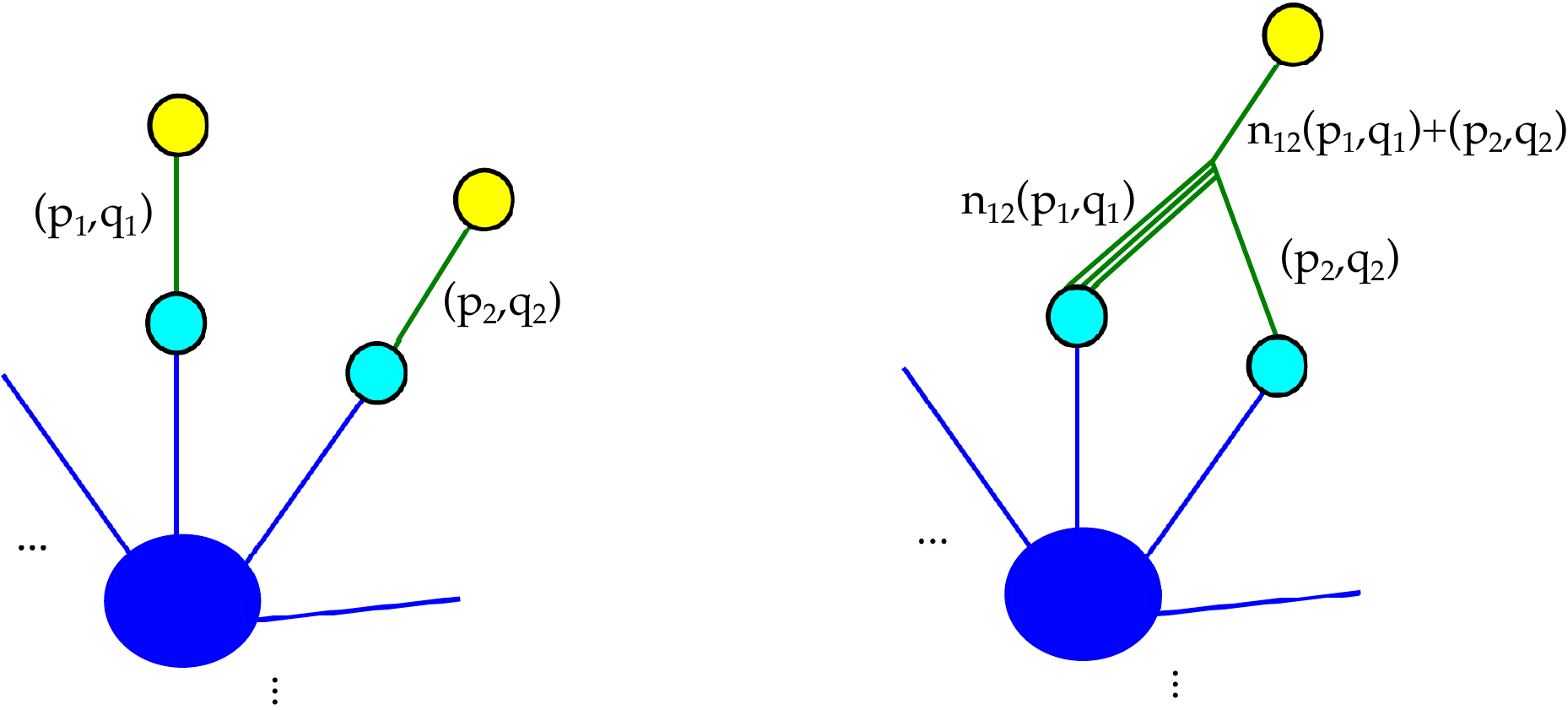}
\caption{On the left hand side is shown a general 5-brane web (blue) indicating the external $(p,q)$ 5-branes, ending on 7-branes (cyan). From these emanate $(p,q)$-strings (green), that end on D3-branes (yellow). Given a pair of external 5-branes, the strings can only form a junction, if they satisfy (\ref{junctioncond}). This is shown on the right hand side. The resulting string can be moved into the brane-web, by moving the D3-brane inside the web, and becomes a local operator. This is the screening of the Wilson lines by local operators, realized in the brane-web.\label{fig:Junct}}
\end{figure}

This observation about filling in of white dots can be understood by considering the Wilson lines in the $(p,q)$-web, which correspond to $(p,q)$-strings, which stretch to infinity (or end on D3-branes at finite distance) \cite{Assel:2018rcw, Uhlemann:2020bek}). A pair of strings ending on 7-branes $(p_1, q_1)$ and $(p_2, q_2)$  can form a single string junction if 
\be
\det \left(\begin{matrix} p_1 &q_1\\ p_2 & q_2 \end{matrix}\right)   = \pm 1 \,.
\ee
Consider a brane web with external 5-branes emanating. Consider two of these of type $(p_1, q_1)$ and $(p_2, q_2)$, which each end on 7-branes at finite distance, of the same $(p,q)$-type. From these 7-branes we can have $(p, q)$ D-strings emanating, which correspond to the Wilson lines (we can end these on D3-branes). Let 
\be
\left| \det \left(\begin{matrix} p_1 &q_1\\ p_2 & q_2 \end{matrix}\right)\right|   =  n_{1,2} \,.
\ee
Then these strings can form a junction satisfying \cite{Bergman:1998ej}
\be\label{junctioncond}
{n_{1,2} (p_1, q_1)} + (p_2, q_2) \qquad \rightarrow \qquad \left((p_2 , q_2) + n_{1,2} (p_1, q_1)\right) \,.
\ee
These can end on D3-branes and can be moved back into the web. This is the analog of the screening of Wilson lines by local operators and is illustrated in figure \ref{fig:Junct}. For a given 5-brane web, each external 5-brane gives rise to Wilson line, in the fashion above. Considering pair-wise the possible junctions determines which Wilson loops are screened. Taking the gcd over these computes the overall screening by all possible string junctions in the web. 
This of course is precisely encoded in the expression (\ref{WebOne}) and the resulting 1-form symmetry. 

From this perspective it is also clear why in a GTP with white dots, the 1-form symmetry is computed from the GTP obtained by filling all white dots and converting the diagram to black dots. A white dot corresponds to two 5-branes ending on the same 7-brane, whereas a black dot along a edge corresponds to two parallel 5-branes ending on one 7-brane each. 
In the former configuration by not including this dot, we would not consider the complete set of strings. The 7-branes are not essential in this, as we can send these to infinity. By not including the white dots, we would not account for all possible strings (Wilson lines), as there can be Wilson lines ending on either of the 5-branes, that end on the 7-brane.

\section*{Acknowledgements}
We thank Fabio Apruzzi, Pietro Benetti Genolini, Antoine Bourget, Cyril Closset, Yi-Nan Wang, Gabi Zafrir and in particular Julius Eckhard for discussions. 
The work of LB is supported by NSF grant PHY-1719924.
The work of SSN is
supported by the ERC Consolidator Grant number 682608 "Higgs bundles: Supersymmetric Gauge Theories and Geometry (HIGGSBNDL)". SSN acknowledges support also from the Simons Foundation.

\bibliographystyle{ytphys}
%\small 
%\baselineskip=.94\baselineskip
\let\bbb\bibitem\def\bibitem{\itemsep4pt\bbb}
\bibliography{ref}

\end{document}